\documentclass[reprint,superscriptaddress,amsmath,amssymb,aps,prx,floatfix]{revtex4-2}
\usepackage{pdfpages,pgffor,graphicx,dcolumn,bm,xr,physics,chemformula,siunitx,color,soul,float,appendix, subfigure, booktabs,array, multirow}

\newcommand{\bmk}{{\bf k}}
\newcommand{\bmq}{{\bf q}}
\newcommand{\bmqq}{{\bf Q}}
\newcommand{\bmr}{{\bf r}}
\newcommand{\bmrr}{{\bf R}}

\renewcommand{\braket}[1]{\expval{#1}}

\newcommand{\ifc}{C_{\kappa \alpha p, \kappa' \alpha' p'} }
\newcommand{\iifc}{C^{-1}_{\kappa \alpha p, \kappa' \alpha' p'} }

\newcommand{\dtau}{\Delta \tau}

\newcommand{\kpa}{{\kappa \alpha p }}
\newcommand{\kpaa}{{\kappa' \alpha' p' }}

\newcommand{\vck}{{vc\bmk }}
\newcommand{\vckk}{{v'c'\bmk' }}

\newcommand{\fourr}{{ \bmr_{\rm e}, \bmr_{\rm h}; \bmr_{\rm e}', \bmr_{\rm h}'}}
\newcommand{\fourrr}{{ \bmr_{\rm e}'', \bmr_{\rm h}''; \bmr_{\rm e}''', \bmr_{\rm h}'''}}
\newcommand{\dfourr}{d\bmr_{\rm e} d\bmr_{\rm h} d\bmr_{\rm e}' d\bmr_{\rm h}'}

\colorlet{blue}{blue!70!black} 
\colorlet{red}{red!60!black}

\makeatletter
\AtBeginDocument{\let\LS@rot\@undefined}
\makeatother

\begin{document}

\raggedbottom

\title{Theory of excitonic polarons:\\[3pt] From models to first-principles calculations}

\author{Zhenbang Dai}
\affiliation{Oden Institute for Computational Engineering and Sciences, The University of Texas at Austin, Austin, Texas 78712, USA}
\affiliation{Department of Physics, The University of Texas at Austin, Austin, Texas 78712, USA}
\author{Chao Lian}
\affiliation{Oden Institute for Computational Engineering and Sciences, The University of Texas at Austin, Austin, Texas 78712, USA}
\affiliation{Department of Physics, The University of Texas at Austin, Austin, Texas 78712, USA}
\author{Jon Lafuente-Bartolome}
\affiliation{Oden Institute for Computational Engineering and Sciences, The University of Texas at Austin, Austin, Texas 78712, USA}
\affiliation{Department of Physics, The University of Texas at Austin, Austin, Texas 78712, USA}
\author{Feliciano Giustino}%
\email{fgiustino@oden.utexas.edu}
\affiliation{Oden Institute for Computational Engineering and Sciences, The University of Texas at Austin, Austin, Texas 78712, USA}
\affiliation{Department of Physics, The University of Texas at Austin, Austin, Texas 78712, USA}
        
\date{\today}

\begin{abstract}
Excitons are neutral excitations that are composed of electrons and holes bound together by their attractive Coulomb interaction. The electron and the hole forming the exciton also interact with the underlying atomic lattice, and this interaction can lead to a trapping potential that favors exciton localization. The quasi-particle thus formed by the exciton and the surrounding lattice distortion is called excitonic polaron. Excitonic polarons have long been thought to exist in a variety of materials, and are often invoked to explain the Stokes shift between the optical absorption edge and the photo-luminescence peak. However, quantitative \textit{ab initio} calculations of these effects are exceedingly rare.  In this manuscript, we present a theory of excitonic polarons that is amenable to first-principles calculations. We first apply this theory to model Hamiltonians for Wannier excitons experiencing Fr\"{o}hlich or Holstein electron-phonon couplings. We find that, in the case of Fr\"{o}hlich interactions, excitonic polarons only form when there is a significant difference between electron and hole effective masses. Then, we apply this theory to calculating excitonic polarons in lithium fluoride \textit{ab initio}. The key advantage of the present approach is that it does not require supercells, therefore it can be used to study a variety of materials hosting either small or large excitonic polarons. This work constitutes the first step toward a complete {\it ab initio} many-body theory of excitonic polarons in real materials.
\end{abstract}
\maketitle


\section{Introduction}
\label{sec:intro}
Excitons are composite quasiparticles formed when an electron and a hole in a crystal bind together under the effect of their attractive Coulomb interaction~\cite{rohlfing2000electron}. These quasiparticles constitute one of the cornerstones of condensed matter physics, as they encode a wealth of information on quantum many-body effects and emergent phenomena in solids~\cite{combescot2015excitons}. Investigations of exciton photophysics range from the coherent manipulation of exotic quantum phases in moiré quasicrystals~\cite{mak2022semiconductor} to Floquet engineering of time crystals~\cite{kobayashi2023floquet}.

\begin{figure}
    \centering
    \includegraphics[width=0.3\textwidth]{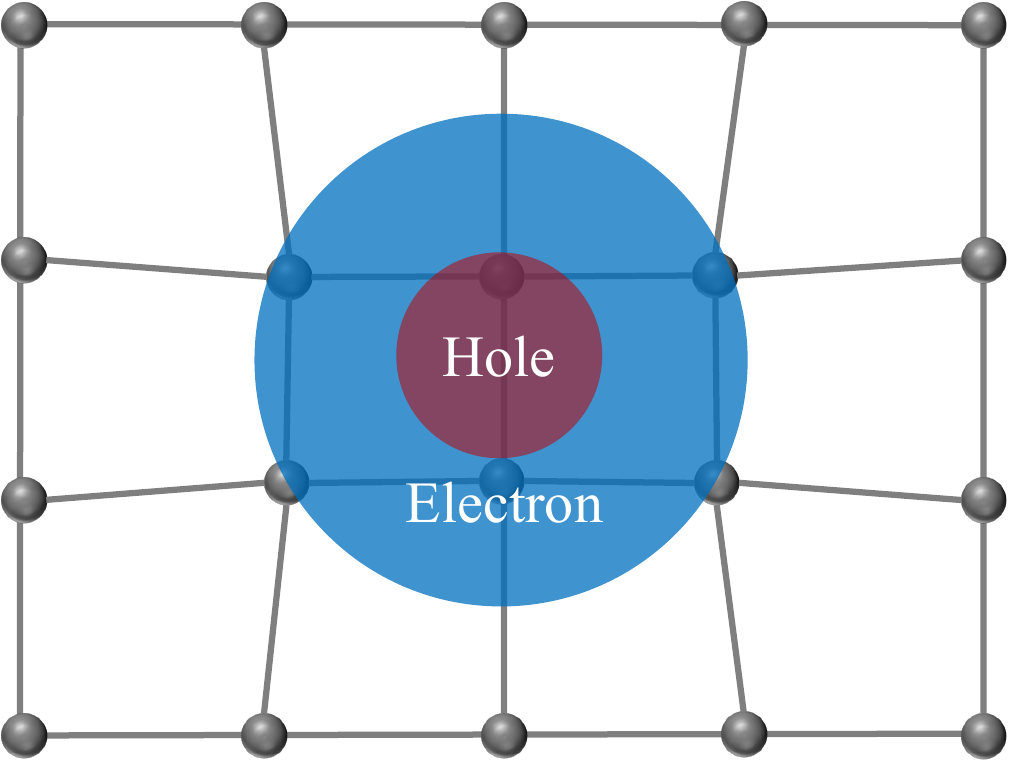}
    \caption{Schematic illustration of an excitonic polaron. The exciton is a neutral excitation, but spatial fluctuations of its net charge density interact electrostatically with the ionic lattice. In some materials, these interactions are sufficiently strong to cause a lattice distortion, which in turn stabilizes the exciton, leading to the formation of an excitonic polaron.}
    \label{fig:explrn_illu}
\end{figure}

In some materials, the spatial fluctuations of the electric charge density of the exciton can polarize the surrounding crystal lattice, and this distortion can promote in turn the spatial localization of the exciton (Fig.~\ref{fig:explrn_illu}). In analogy to charged polarons, where an electron or hole will be self-localized via induced lattice distortions~\cite{devreese2010frohlich, sio2019polarons}, the new type of quasiparticle formed by the feedback loop between exciton and crystal lattice is called {\it excitonic polaron}~\cite{iadonisi1987polaronic, iadonisi1984electron,iadonisi1983excitonic}. 
Intuitively, an excitonic polaron can be understood as as an exciton accompanied by a phonon cloud. 
In materials where the interaction between excitons and phonons is very strong, the same mechanism leads to the emergence of self-trapped excitons, which can be understood as intrinsic defect-like excited-states of an otherwise perfect lattice~\cite{williams1990self, fowler1973theory, ismail2005self}. 
Self-trapped excitons in solids exhibit intriguing optical properties, such as distinctive vibronic lineshapes that are typically observed in molecular chromophores and light-harvesting complexes~\cite{williams1990self, ismail2005self, abfalterer2020colloidal,guizard1996time}.
{We note that the names ``excitonic polarons'' and ``self-trapped excitons'' refer to the same physical mechanism of polaronic stabilization of the exciton. 
The difference between these two concepts resides in the strength of the electron-phonon coupling, the resulting phonon-induced localization of the exciton wavefunction, and the magnitude of the hopping barrier for exciton migration. 
Therefore, in this article, we use the two naming conventions interchangeably. }

Unlike polarons, which can be described to a good approximation within a single-particle picture~\cite{sio2019ab, sio2023polarons, lafuente2022ab, lafuente2022unified}, excitons are inherently of many-body character, thus making the theoretical description of excitonic polarons much more challenging than for polarons. 

In early work on excitonic polarons, Iadonisi and Bassani considered the Wannier exciton model and Fr\"{o}hlich electron-phonon interaction~\cite{iadonisi1983excitonic}. They wrote the Hamiltonian of the excitonic polaron by including the kinetic energies of electrons and holes, their mutual Coulomb attraction, their respective interaction with phonons, and the energy of bare phonons. Within this model, they discussed solutions based on a trial wave function with parameters optimized variationally. This work contributed to shaping the key conceptual aspects of the physics of excitonic polarons; however, since the dispersions of the exciton bands were not taken into account, this formalism is biased to find localized solutions.

In the area of {\it ab initio} calculations, excitonic polarons have been investigated using the $\Delta$SCF (self-consistent field) method to obtain excited-state forces~\cite{mackrodt2022self,du2019electronic}. The limitation of this approach is that it misses electron-hole correlations in the excitonic state~\cite{onida2002electronic, rohlfing2000electron}. Coupled-cluster calculations have also been reported, namely for the self-trapped excitons in quartz~\cite{van2003ab}, but the high computational cost limits the size of the clusters that can be investigated, and this leads to numerical uncertainty due to convergence issues.
The first attempt to compute self-trapped excitons within a many-body Green's function framework was reported by Ismail-Beigi and Louie~\cite{ismail2005self}, who succeeded to compute forces in the excited state by differentiating the Bethe-Salpeter equation (BSE) Hamiltonian with respect to the atomic displacements. 
This method was demonstrated for the self-trapped triplet exciton in quartz. 
Depite being conceptually elegant and numerically accurate, this approach requires performing BSE calculations in large supercells for all but the smallest excitonic polarons, making the computational cost prohibitive in many systems of interest.

In previous work, some of us demonstrated that it is possible to compute small and large polarons using the Kohn-Sham band structures, phonon dispersions, and electron-phonon matrix elements, and without using supercells~\cite{sio2019ab,sio2023polarons,sio2019polarons}. The key concept of that approach is that the wavefunction and atomic displacements of the polaron can be expressed as linear superpositions of Kohn-Sham states and normal modes obtained from unit-cell calculations. In this manuscript, we expand on this idea by tackling the calculation of excitonic polarons and self-trapped excitons using a combination of the BSE approach and linear-response calculations of exciton-phonon couplings~\cite{chen2020exciton, antonius2022theory}. Our aim is to provide a formalism and computational method to compute excitonic polarons using solely information calculated in the crystal unit cell. A short summary of this methodology is reported in the companion manuscript~\cite{dai2023explrn_prl}.

The manuscript is organized as follows: In Sec.~\ref{sec:theory}, we derive the main formalism to compute excitonic polarons, and establish the connection with BSE calculations and with the calculation of electron and hole polarons. We also present two alternative formulations of the theory, which are most useful in the context of model Hamiltonians and \textit{ab initio} calculations, respectively. In Sec.~\ref{sec:model} we apply the present theory to model Hamiltonians in order to analyze the qualitative aspects of the solutions in simple cases. In particular, we discuss the Wannier exciton in the presence of Fr\"ohlich or Holstein electron-phonon couplings. Section~\ref{sec:abinitio} presents first-principles calculations of excitonic polarons for LiF as a first application of this methodology. Here, we also discuss the gauge invariance of the theory, implementation details, and convergence tests. In Sec.~\ref{sec:summary}, we draw our conclusions and propose possible avenues for future work.

Throughout this manuscript, ``electron-phonon coupling'' will generally denote the interaction between a phonon and an electronic state, either occupied or empty. However, in those cases where we need to differentiate between valence and conduction bands, we will use ``hole-phonon coupling'' or ``electron-phonon coupling'', respectively. 

\section{Ab initio Theory of Excitonic Polarons}
\label{sec:theory}

We begin our derivation by expressing the total energy of a distorted lattice in a neutral excited state as the sum of its total energy in the electronic ground-state and the BSE excitation energy, following Ref.~\citenum{ismail2003excited}:
\begin{align}
    &E\left[ \Psi(\bmr_{\rm e},\bmr_{\rm h}),
    \{ \Delta \tau_{\kpa} \}
    \right]
    =E_{\mathrm{DFT}}\left[ \{ \Delta \tau_{\kpa} \} \right]
    \nonumber \\
    &+\int_{\rm sc}
    \Psi^*(\bmr_{\rm e},\bmr_{\rm h}) H_{\mathrm{BSE}}(\fourr) \Psi(\bmr_{\rm e}',\bmr_{\rm h}')
    d\bmr.\label{eq.1}
\end{align}
In this expression, we use the subscript ``DFT'' to indicate that the total energy of the electronic ground state is computed at the level density functional theory (DFT). The formalism remains unchanged if this total energy is computed using more accurate techniques. In Eq.~\eqref{eq.1}, $H_{\mathrm{BSE}}$ is the BSE Hamiltonian for the distorted structure~\cite{rohlfing1998electron, rohlfing2000electron}; $\Psi(\bmr_{\rm e},\bmr_{\rm h})$ is the exciton wavefunction, with 
$\bmr_{\rm e}$ and $\bmr_{\rm h}$ denoting the electron and hole coordinates, respectively. 
The integral extends over the Born-von K\'arm\'an (BvK) supercell, and the integration variable $d\bmr$ is a short-hand notation for $\dfourr$. For small atomic displacements, $E_{\mathrm{DFT}}$ can be expressed in terms of the matrix of interatomic force constants $\ifc$, 
\begin{align}\label{eq.2}
    &E_{\mathrm{DFT}}\left[ \{ \Delta \tau_{\kpa} \} \right]
    \nonumber \\
    =& E_0 + 
    \frac{1}{2} \sum_{\substack{\kpa\\ \kpaa}} \ifc \Delta \tau_{\kpa} \Delta \tau_{\kpaa},
\end{align}
where $E_0$ denotes the ground-state energy of the undistorted structure, and $\Delta \tau_{\kpa}$ is the displacement of the atom $\kappa$ in the unit cell $p$ along the Cartesian direction $\alpha$, with respect to the undistorted structure. 
The second term in this equation is the elastic energy associated with the lattice distortion. As in Ref.~\onlinecite{sio2019polarons}, the energy is truncated to the second order in the displacements; {despite the neglect of anharmonicity, which is included in previous works based on supercells \cite{lafuente2022ab, lafuente2022unified}}, this approximation proved successful in calculations of both small and large polarons, comparing well with direct hybrid-functional calculations~\cite{sio2019polarons, sio2019ab, sio2023polarons, lafuente2022ab, lafuente2022unified}. 
{Note that, in this work, we describe nuclei within the adiabatic and classical approximation, as in DFT calculations. 
Strictly speaking, this choice makes our formalism best suited to described the strong coupling limit, as in the Pekar polaron model~\cite{lafuente2022ab}.}
Combining Eqs.~\eqref{eq.1} and \eqref{eq.2}, we can write the total energy as:
\begin{eqnarray}
    \label{eqn:etot_rs}
    &&E\left[ \Psi(\bmr_{\rm e},\bmr_{\rm h}) ,
    \{ \Delta \tau_{\kpa} \}
    \right]
    = E_0\nonumber \\
    &&
    \qquad +\int_{\rm sc}
    \Psi^*(\bmr_{\rm e},\bmr_{\rm h}) H_{\mathrm{BSE}}(\fourr) \Psi(\bmr_{\rm e}',\bmr_{\rm h}')
    d\bmr
    \nonumber \\
    &&\hspace{20pt}+
    \,\frac{1}{2}\!\! \sum_{\substack{\kpa\\ \kpaa}} \ifc \Delta \tau_{\kpa} \Delta \tau_{\kpaa}.
\end{eqnarray}
To obtain excitonic polarons, we require that the exciton wavefunction and the atomic displacements minimize the total energy in Eq.~\eqref{eqn:etot_rs}. We use the method of Lagrange multipliers, and set to zero the functional derivatives of $E\left[ \Psi(\bmr_{\rm e},\bmr_{\rm h}) , \{ \Delta \tau_{\kpa} \} \right]$ with respect to $\Psi(\bmr_{\rm e},\bmr_{\rm h})$ and $\{ \Delta \tau_{\kpa} \}$, subject to the normalization constraint $\int_{\rm sc}\abs{\Psi(\bmr_{\rm e},\bmr_{\rm h})}^2 d\bmr_{\rm e} d\bmr_{\rm h}=1$.
By expanding $H_{\mathrm{BSE}}$ up to linear order in $\Delta \tau_\kpa$ and after some straightfoward algebra, we arrive at the following coupled nonlinear eigenvalue problem:
\begin{eqnarray}
    \label{eqn:minimize_psi}
    &&\int_{\rm sc}\!\!
    H^0_{\mathrm{BSE}}(\fourr) \,\Psi(\bmr_{\rm e}',\bmr_{\rm h}')\,d\bmr_{\rm e}' d\bmr_{\rm h}' 
    \nonumber \\
    &+&
    \sum_{\kpa}
    \int_{\rm sc}
    \frac{\partial \hat{H}^0_{\mathrm{BSE}}(\fourr)}{\partial \tau_{\kpa}}\,
    \Psi(\bmr_{\rm e}',\bmr_{\rm h}')\,
    d\bmr_{\rm e}' d\bmr_{\rm h}'\, \dtau_{\kpa}
    \nonumber \\
    &=& \varepsilon \,\Psi(\bmr_{\rm e},\bmr_{\rm h}), \\
    \label{eqn:minimize_tau}
    &&\hspace{-15pt}\dtau_{\kpa} 
    = 
    -\!\!\sum_{\kappa'\alpha' p'}\iifc
    \nonumber \\
    &&\times\int_{sc} \!\!
    \Psi^*(\bmr_{\rm e},\bmr_{\rm h})  
    \frac{\partial \hat{H}^0_{\mathrm{BSE}}(\fourr)}{\partial \tau_{\kpaa}}
    \Psi(\bmr_{\rm e}',\bmr_{\rm h}')\,d\bmr,
\end{eqnarray}
where $H^0_{\mathrm{BSE}}$ is the BSE Hamiltonian for the undistorted system, and the eigenvalue $\varepsilon$ is the Lagrange multiplier.

Alternatively, the set of equations Eqs.~(\ref{eqn:minimize_psi}) and (\ref{eqn:minimize_tau}) can formally be combined in a single nonlinear problem for the exciton wavefunction:
\begin{align}
    \label{eqn_rs:polaron_eqn}
    &\int_{sc} d\bmr_e' d\bmr_h'
    \Bigg[
    \hat{H}^0_{\mathrm{BSE}}(\fourr) 
    \nonumber \\
    &-
    \int_{sc} d\bmr'
    \Psi^*(\bmr_e'',\bmr_h'')
    K(\{ \bmr \})
    \Psi(\bmr_e''',\bmr_h''')
    \Bigg]
    \Psi(\bmr_e',\bmr_h') 
    \nonumber \\
    =& \varepsilon \Psi(\bmr_e,\bmr_h),
\end{align}
with the kernel $K(\{ \bmr \})$ being defined as:
\begin{align}
    \label{eqn_rs:kernel}
    K(\{ \bmr \})
    =&
    \sum_{\substack{\kpa \\ \kpaa}}
    \frac{\partial \hat{H}^0_{\mathrm{BSE}}(\fourr)}{\partial \tau_{\kpa}}
    \nonumber \\
    &\times
    \iifc
    \frac{\partial \hat{H}^0_{\mathrm{BSE}}(\fourrr)}{\partial \tau_{\kpaa}},
\end{align}
and $\{ \bmr \}$ stands for $(\fourr; \fourrr)$.

\subsection{Transition basis approach}
\label{sec:transition}

By solving Eq.~(\ref{eqn_rs:polaron_eqn}), in principle we can obtain the wave function $\Psi(\bmr_e,\bmr_h)$ of the excitonic polaron. However, in this form, a large BvK supercell is still needed in order to describe this quasiparticle. To circumvent this difficulty, we follow the standard approach employed for solving the BSE equations: we expand $\Psi(\bmr_e,\bmr_h)$ in a transition basis within the Tamm-Dancoff approximation~\cite{onida2002electronic, rohlfing2000electron}:
\begin{align}
    \label{eqn_rs:basis0}
    \Psi(\bmr_e, \bmr_h) = \frac{1}{\sqrt{N_p}}\sum_{vc} \tilde{A}_{vc} \psi^0_c (\bmr_e) \psi^{0*}_v (\bmr_h).
\end{align}
In this expression, $\psi^0_n$ denotes Kohn-Sham eigenstates of the undistorted structure, and $N_p$ is the number of primitive cells in BvK supercell. The subscripts $v$ and $c$ denote valence and conduction states, respectively. For notation brevity, we temporarily suppress the dependence of all quantities on the crystal momentum; we will restore the momentum in the final equations. Using Eq. \eqref{eqn_rs:basis0}, we rewrite Eq.~(\ref{eqn_rs:polaron_eqn}) as:
\begin{widetext}
\begin{align}
    \label{eqn_rs:polaron_eqn_2}
    &\sum_{v'c'} \Bigg[\bra{vc} \hat{H}^0_{BSE} \ket{v'c'}
    -\sum_{\substack{v''c'' \\v'''c'''}} 
    \sum_{\substack{\kpa \\ \kpaa}}
    \tilde{A}^*_{v''c''} \tilde{A}_{v'''c'''} 
    C^{-1}_{\kpa,\kpaa}
    \braket{vc \left| \frac{ \partial \hat{H}^0_{BSE}}{\partial \tau_{\kpa}} \right| v'c'}
    \braket{v''c'' \left| \frac{ \partial \hat{H}^0_{BSE}}{\partial \tau_{\kpaa}} \right| v'''c'''}
    \Bigg] \tilde{A}_{v'c'}
    \nonumber \\
    =&\varepsilon \tilde{A}_{vc}.
\end{align}
\end{widetext}
In the above expression, brakets have the following meaning:
\begin{align}
    &\bra{vc} \hat{{O}} \ket{v'c'}
    \nonumber \\
    =&
    \int_{sc} 
    d\bmr
    \psi^{0*}_c(\bmr_e)\psi^{0}_v(\bmr_h)
    \hat{{O}}(\fourr) 
    \psi^0_{c'} (\bmr_e') \psi^{0*}_{v'} (\bmr_h'),
\end{align}
where $\hat{{O}}$ is an operator that depends on the electron and hole coordinates.
The first term of Eq.~(\ref{eqn_rs:polaron_eqn_2}) corresponds to the matrix elemens of the BSE Hamiltonian in the undistrorted structure, and can be computed using any package that implements the BSE method~\cite{deslippe2012berkeleygw,marini2009yambo}. To evaluate the second term in the square brackets, we rewrite the matrix elements in the sum follows:
\begin{align}
    \label{eqn_rs:derivative_h}
    &\braket{vc \left| \frac{ \partial \hat{H}^0_{BSE}}{\partial \tau_{\kpa}} \right| v'c'}
    \nonumber \\
    =&-\braket{(\partial_{\tau}v)c \left| \hat{H}^0_{\mathrm{BSE}}  \right| v'c'}
    -\braket{v(\partial_{\tau}c) \left| \hat{H}^0_{\mathrm{BSE}}   \right| v'c'}
    \nonumber \\
    &
    -\braket{vc \left| \hat{H}^0_{\mathrm{BSE}}  \right| (\partial_{\tau}v')c'}
    -\braket{vc \left| \hat{H}^0_{\mathrm{BSE}}  \right| v'(\partial_{\tau}c')}
    \nonumber \\
    &+\partial_\tau
    \braket{vc \left| \hat{H}^0_{\mathrm{BSE}} \right| v'c'},
\end{align}
with $\partial_\tau$ being understood as ${\partial}/\partial_{\tau_{\kpa}}$. The first four terms on the right-hand-side share the same structure, therefore we focus on the first term as a representative case. To recast this term in a manageable form, we express the variation of the Kohn-Sham wavefunctions using first-order perturbation theory~\cite{giustino2017electron}, and we use the definition of the BSE Hamiltonian $\braket{vc \left| \hat{H}^0_{BSE} \right| v'c'}=(\epsilon_c - \epsilon_v)\delta_{cc'}\delta_{vv'}+\braket{ vc \left| \hat{K}_{\mathrm{BSE}}^0 \right| v'c'}$, where $\epsilon_n$ denote quasi-particle energies of the undistorted structure and $\hat{K}_{\mathrm{BSE}}^0$ is the BSE kernel~\cite{rohlfing1998electron}. We find:
\begin{align}
    \label{eqn:first_term_v2}
    &\braket{(\partial_{\tau}v)c \left| \hat{H}^0_{\mathrm{BSE}}  \right| v'c'}
    \nonumber \\
    =&
    \frac{\braket{v'|\partial_{\tau} \hat{V}_{\mathrm{SCF}}|v}}{\epsilon_v - \epsilon_v'}
    (\epsilon_c - \epsilon_v') \delta_{cc'} 
    +\braket{ (\partial_{\tau}v)c | \hat{K}_{\mathrm{BSE}}^0 | v'c'}.
\end{align}
Now we consider the fifth term on the right-hand side of Eq.~(\ref{eqn_rs:derivative_h}),
\begin{align}
    \label{eqn:fifth_term}
    &\partial_\tau\braket{vc \left| \hat{H}^0_{\mathrm{BSE}} \right| v'c'}
    \nonumber \\
    =&{\partial_\tau} (\epsilon_c - \epsilon_v) \delta_{cc'} \delta_{vv'}
    +
    \partial_\tau \braket{ vc | \hat{K}_{\mathrm{BSE}}^0 | v'c'}.
\end{align}
and we combine Eqs.~(\ref{eqn:first_term_v2}) and (\ref{eqn:fifth_term}). By neglecting the change of the BSE kernel upon lattice distortion as in Refs.~\citenum{ismail2003excited, chen2020exciton, antonius2022theory}, we obtain:
\begin{align}
    \label{eqn_rs:derivative_h_final}
    &\braket{vc \left| \frac{ \partial \hat{H}^0_{\mathrm{BSE}}}{\partial \tau_{\kpa}} \right| v'c'}
    \nonumber \\
    =&
    -\braket{v' \left| \frac{\partial \hat{V}_{\mathrm{SCF}}}{\partial \tau_{\kpa}} \right| v} \delta_{cc'}
    +\braket{c' \left| \frac{\partial \hat{V}_{\mathrm{SCF}}}{\partial \tau_{\kpa}} \right|  c}^* \delta_{vv'}.
\end{align}
The final set of equations is obtained by restoring the dependence of all quantities on the crystal momentum:
\begin{align}
    \label{eqn:expansion_transition}
    \Psi(\bmr_e, \bmr_h) = \frac{1}{\sqrt{N_p}} \sum_{vc\bmk\bmqq} \tilde{A}^{\bmqq}_{\vck} \psi^0_{c\bmk+\bmqq} (\bmr_e) \psi^{0*}_{v\bmk} (\bmr_h),
\end{align}
where $\bmk$ is the crystal momentum of the electron or hole, and $\bmqq$ is the crystal momentum of the exciton. Since we require the normalization condition $\int_{\mathrm{sc}}\abs{\psi_{n\bmk}(\bmr)}^2 d\bmr = 1$, we have the sum rule $N_p^{-1}\sum_{\vck\bmqq} \abs{\tilde{A}_{\vck}^{\bmqq}}^2 = 1$. {In Eq.~\eqref{eqn:expansion_transition} and in the following, summations over crystal momenta run over uniform Brillouin-zone grids with $N_p$ points.
We employ the following standard relations for the matrix of interatomic force constants and the electron-phonon coupling matrix elements~\cite{sio2019ab}:
\begin{align}
    ({C}^{-1})_{\kappa \alpha p, \kappa' \alpha' p' }
    &=
    \frac{1}{N}\sum_{\bmq \nu} \frac{e_{\kappa \alpha, \nu}(\bmq)e^*_{\kappa' \alpha', \nu}(\bmq)}
    {\sqrt{M_{\kappa} M_{\kappa'}} \omega^2_{\bmq \nu}} 
    e^{i\bmq \cdot(\bmrr_p - \bmrr_{p'})},
    \label{eqn:iifc}
    \\
    g_{mn\nu}(\bmk,\bmq)
    =&\sum_{\kpa}
    \sqrt{\frac{\hbar}{2M_{\kappa}\omega_{\bmq \nu}}}
    e_{\kappa\alpha,\nu}(\bmq)e^{i\bmq \cdot \bmrr_p}
    \nonumber \\
    &\times
    \braket{\psi_{m\bmk+\bmq}^0\left|\frac{\partial\hat{V}_{\mathrm{SCF} }}{\partial{\tau_{\kpa}} }\right| \psi_{n\bmk}^0}, \label{eqn:def_eph}
\end{align}
where $M_{\kappa}$ is the mass of atom $\kappa$; $e_{\kappa \alpha, \nu}(\bmq)$ is the polarization vector of the phonon with momentum $\bmq$, branch $\nu$, and frequency $\omega_{\bmq\nu}$; $\bmrr_p$ is the lattice vector of the $p$-th unit cell in the BvK supercell~\cite{giustino2017electron}. Using Eqs.~\eqref{eqn_rs:derivative_h_final}-\eqref{eqn:def_eph}, Eq.~(\ref{eqn_rs:polaron_eqn_2}) can be rewritten as follows:
\begin{align}
\sum_{ \substack{ \vckk \\ \bmqq'}} &\tilde{A}^{\bmqq'}_{\vckk}
\bigg[ (H^0_{\mathrm{BSE}})_{v\bmk c\bmk+\bm{Q},v'\bmk' c'\bmk'+\bm{Q}'}
\delta_{\bm{Q},\bm{Q}'} 
\nonumber \\
&-
\frac{2}{N_p}\sum_{\bmq \nu}
\tilde{B}_{\nu} \tilde{\mathcal{G}}^{\bmq\nu}_{\vck,\vckk} (\bmqq,\bmqq',\bmq)
\bigg] 
= \varepsilon \tilde{A}^{\bmqq}_{\vck}, \label{eqn_rs:polaron_eqn_final}
\end{align}
\begin{align}
\tilde{B}_{\bmq \nu} 
=&
\frac{1}{N_p \hbar \omega_{\bmq \nu}} 
\sum_{\substack{\vck \\ \bmqq}} 
\tilde{A}^{\bm{Q}*}_{\vck}
\Bigg[
\sum_{c'}
\tilde{A}^{\bm{Q+\bmq}}_{vc'\bmk}
g_{cc'\nu}(\bmk + \bmqq +\bmq,-\bmq)
\nonumber \\
&-
\sum_{v'}
\tilde{A}^{\bm{Q+q}}_{v'c\bmk-\bmq}
g_{v'v\nu}(\bmk, -\bmq)    
\Bigg], \label{eqn_rs:b_transition}
\end{align}
\begin{align}
\tilde{\mathcal{G}}^{\nu}_{\vck,\vckk} (\bmqq,\bmqq',\bmq)
=&
g_{cc'\nu} (\bmk'+\bmqq',\bmq) \delta_{\bmq, \bmqq-\bmqq'}
\delta_{vv'}
\delta_{\bmk,\bmk'}
\nonumber \\
&-
g_{v'v\nu} (\bmk,\bmq) 
\delta_{\bmq, \bmk'-\bmk}
\delta_{cc'}
\delta_{\bmq,\bmqq-\bmqq'}, \label{eqn_rs:g_transition}
\end{align}
Equation~\eqref{eqn_rs:polaron_eqn_final} defines an eigenvalue problem for the coefficients $\tilde{A}_{\vck}^\bmqq$ which make up the excitonic polaron wavefunction. The matrix to be diagonalized depends on the atomic displacements via the coefficients $\tilde{B}_{\bmq \nu}$ given by Eq.~\eqref{eqn_rs:b_transition}, and the $\tilde{A}_{\vck}^\bmqq$ and $\tilde{B}_{\bmq \nu}$ coefficients are coupled by the coupling matrix elements $\tilde{\mathcal{G}}^{\nu}_{\vck,\vckk} (\bmqq,\bmqq',\bmq)$ given in Eq.~\eqref{eqn_rs:g_transition}. The ingredients required to solve Eqs.~\eqref{eqn_rs:polaron_eqn_final}-\eqref{eqn_rs:g_transition} can be obtained from existing packages that implement the BSE method and packages that calculate electron-phonon couplings.

Equations~\eqref{eqn_rs:polaron_eqn_final}-\eqref{eqn_rs:g_transition} are most convenient to investigate excitonic polarons within model Hamiltonians, as we show in Sec.~\ref{sec:etot_transition} for the Wannier exciton model with Fr\"{o}hlich electron-phonon interactions.
%
%
However, these equations are not optimal for \textit{ab initio} calculations, because the dimension of the coefficients $\tilde{A}_{\vck}^\bmqq$ is $N_\bmk \times N_\bmqq \times N_v \times N_c$, where $N_\bmk$ is the number of electron and phonon crystal momenta, $N_\bmqq$ is the number of exciton crystal momenta, $N_v$ is the number of valence bands, and $N_c$ is the number of conduction bands. This scaling can be prohibitive even for relatively coarse Brillouin-zone grids. The origin of this problematic scaling is that Eqs.~\eqref{eqn_rs:polaron_eqn_final}-\eqref{eqn_rs:g_transition} are designed to construct excitonic polarons starting from Kohn-Sham electron and hole wavefunctions, therefore they accomplish two goals simultaneously: (i) to describe exciton formation out of non-interacting electron-hole pairs, and (ii) to describe phonon-induced localization of these excitons. An alternative and computationally more convenient strategy would be to first build excitonic states, and then to incorporate their interactions with the atomic lattice. This alternative strategy, which we call the ``exciton basis approach'', is described in the next section. 

\subsection{Exciton basis approach}
\label{sec:exciton}

The heavy computational cost related to Eqs.~(\ref{eqn_rs:polaron_eqn_final})-(\ref{eqn_rs:g_transition}) can effectively be avoided by performing a basis set transformation, from the transition basis of Eq.~\eqref{eqn_rs:basis0} to the following exciton basis:
\begin{align}
    \label{eqn:basis_transform}
    \Omega_{s\bmqq}(\bmr_e,\bmr_h) 
    =
    \sum_{\vck} a_{\vck}^{s\bmqq} 
    \psi_{c\bmk+\bmqq}(\bmr_e) \psi^*_{v\bmk}(\bmr_h),
\end{align}
with the inverse transform:
\begin{align}
    \label{eqn:basis_transform_rev}
    \psi_{c\bmk+\bmqq}(\bmr_e) \psi^*_{v\bmk}(\bmr_h)
    =
    \sum_{s} a_{\vck}^{s\bmqq*} 
    \Omega_{s\bmqq}(\bmr_e,\bmr_h).
\end{align}
In these expressions, $\Omega_{s\bmqq}(\bmr_e, \bmr_h)$ denotes exciton states of the undistorted structure, $a_{\vck}^{s\bmqq}$ are the BSE eigenvectors for the undistorted structure, $s$ is the index of the exciton bands, and $\bmqq$ is the exciton momentum. The coefficients $a_{\vck}^{s\bmqq}$ fulfill the normalization condition $\sum_{\vck} \abs{a_{\vck}^{s\bmqq}}^2=1$. In writing Eq.~(\ref{eqn:basis_transform}), we made use of the Tamn-Dancoff approximation, so that for each exciton momentum $\bmqq$, the number of exciton states equals the number of electron-hole pairs in the BvK supercell~\cite{onida2002electronic, rohlfing2000electron}.

Using the transformation given by Eq.~\eqref{eqn:basis_transform}, the excitonic polaron state can be expressed as a linear superposition of excitons in the undistorted structure:
\begin{align}\label{eq.exbasis}
    \Psi(\bmr_e,\bmr_h) = \sum_{s\bmqq} A_{s\bmqq} \Omega_{s\bmqq}(\bmr_e,\bmr_h).
\end{align}
Accordingly, the coefficients $A_{s\bmqq}$ can be understood as the contributions of each excitonic state (of the undistorted structure) to the excitonic polaron. In this representation, exciton self-localization and breaking of translational symmetry are possible because we allow for coherent superpositions of delocalized exciton states with finite momenta.

To see how the basis transformation in Eq.~\eqref{eqn:basis_transform} reduces the computational cost, we define $A_{s\bmqq}= \sum_{\vck} \tilde{A}_{\vck}^{\bmqq} a_{\vck}^{s\bmqq*}$, and substitute Eq.~(\ref{eqn:basis_transform_rev}) into Eqs.~(\ref{eqn_rs:polaron_eqn_final})-(\ref{eqn_rs:g_transition}). After some algebraic manipulations we find:
\begin{eqnarray}
    \label{eqn:explrneqn}
    &&\sum_{s'\bmqq'} 
    \bigg[ 
    E^0_{s\bmqq}
    \delta_{ss'}
    \delta_{\bm{Q} \bm{Q}'} 
    -\frac{2}{N_p}\sum_{\nu}
    B_{\bmqq-\bmqq' \nu} \mathcal{G}_{ss'\nu}(\bmqq',\bmqq-\bmqq')
    \bigg] 
    \nonumber \\ && \hspace{20pt}\times A_{s'\bmqq'}
    = \varepsilon A_{s\bmqq},  \\
    &&B_{\bmqq \nu}  
    = 
    \frac{1}{N_p \hbar \omega_{\bmqq \nu}} 
    \sum_{\substack{ss'\\ \bmqq'}}
    A_{s'\bmqq'}^* A_{s\bmqq'+\bmqq}
 \mathcal{G}^*_{ss'\nu}( \bmqq', \bmqq), \nonumber \\[-10pt] \label{eqn:bmat}
\end{eqnarray}
where $E_{s\bmqq}^0$ are the BSE eigenvalues of the undistorted structure, and $\mathcal{G}_{ss'\nu}( \bmqq', \bmqq)$ denote exciton-phonon coupling matrix elements as in Refs.~\citenum{chen2020exciton, antonius2022theory}:
\begin{align}
    \label{eqn:exphg}
    \mathcal{G}_{ss'\nu}(\bmqq,\bmq)
    =&
    \sum_{\vck} a_{\vck}^{s\bmqq+\bmq*}  
    \Bigg[
    \sum_{c'}
    g_{cc'\nu} (\bmk+\bmqq,\bmq)
    a_{vc'\bmk}^{s'\bmqq}
    \nonumber \\
    &-\sum_{v'}
    g_{v'v\nu} (\bmk,\bmq) 
    a_{v'c\bmk + \bmq}^{s'\bmqq}
    \Bigg].
\end{align}
{From Eq.~(\ref{eqn:exphg}) we see that the three grids for crystal momentum $\bmk$, phonon wavevector $\bmq$, and exciton center-of-mass momentum $\bmqq$ should be commensurate. 
Moreover, the $\bmk$-grid should be equal to or denser than the $\bmqq$-grid, while the $\bmqq$-grid should be equal to or denser than the $\bmq$-grid.
For simplicity, in this work we use the same grid for $\bmk, \bmq$ and $\bmqq$ sampling in all first principles calculations.
}
For later reference, we call the quantity in the square bracket of Eq.~(\ref{eqn:explrneqn}) as the ``excitonic polaron Hamiltonian''.

As compared with Eqs.~(\ref{eqn_rs:polaron_eqn_final})-(\ref{eqn_rs:g_transition}), Eqs.~(\ref{eqn:explrneqn})-(\ref{eqn:exphg}) can greatly reduce the cost of computing excitonic polarons. In fact, the dimension of the solution vectors of the eigenvalue problem in Eq.~\eqref{eqn:explrneqn} is $N_s \times N_{\bmqq}$, 
where $N_s$ is the number of {included exciton bands, which can be much smaller than $N_c \times N_v \times N_k$ if only a small number of low-energy exciton bands contribute most significantly to the excitonic polarons.
In practical calculations, convergence tests with respect to the number of exciton bands must always be performed.}
This size is thus orders of magnitude smaller than in the transition basis ($N_\bmk \times N_\bmqq \times N_v \times N_c$, cf.~Sec.~\ref{sec:transition}), making the problem tractable in \textit{ab initio} calculations. Also in this case, the ingredients required to solve Eqs.~(\ref{eqn:explrneqn})-(\ref{eqn:exphg}) are the BSE solutions for the undistorted structure and the electron-phonon matrix elements. Both sets of quantities are evaluated using unit-cell calculations, and no supercells are required.

Equations~(\ref{eqn:explrneqn})-(\ref{eqn:exphg}) also allow us to make the formal connection with the previously-developed \textit{ab initio} polaron equations~\cite{sio2019ab}. In fact, if we formally replace the BSE Hamiltonian by the Kohn-Sham Hamiltonian, and the exciton-phonon coupling matrix elements $\mathcal{G}_{ss'\nu}(\bmqq,\bmq)$ by the electron-phonon coupling matrix elements $g_{mn\nu}(\bmk,\bmq)$, Eqs.~(\ref{eqn:explrneqn})-(\ref{eqn:exphg}) reduce precisely to the polaron equations of Ref.~\onlinecite{sio2019ab}.
{Finally, we would like to emphasize again that our approach does not need supercells in real space.  This is because by using a uniform grid of, for example, $N\times N\times N$ wavevectors in the Brillouin zone, we will have Kohn-Sham states, vibrational eigenmodes, and excitons that are defined on an equivalent Born-von-K\'arm\'an supercell of size $N\times N\times N$ primitive cells. 
Since all our calculations (except for the final visualization) are carried out in reciprocal space, there is no need for explicit supercells.}

\subsection{Total energy and displacement pattern}
\label{sec:etot_pattern}

The present formalism allows us to compute the total energy of the excitonic polaron $E_{\mathrm{xp}}$.
To do so, we first substitute Eq.~(\ref{eqn:minimize_tau}) into Eq.~(\ref{eqn:etot_rs}), and then we write the total energy in the transition basis with the help of Eqs.~(\ref{eqn:iifc}) and (\ref{eqn:def_eph}):
\begin{align}
    \label{eqn:transition_etot}
    E_{\mathrm{xp}}
    =&
    \frac{1}{N_p}
    \sum_{\substack{\vck\\ \vckk}}
    \sum_{\bm{Q},\bm{Q}'}
    \tilde{A}^{\bm{Q}*}_{\vck} \tilde{A}^{\bm{Q'}}_{\vckk} 
    \nonumber \\
    &\times 
    \Bigg[ (H^0_{\mathrm{BSE}})_{v\bmk c\bmk+\bm{Q},v'\bmk' c'\bmk'+\bm{Q}'}
    \delta_{\bm{Q},\bm{Q}'} 
    \nonumber \\
    &-\frac{1}{N_p}
    \sum_{\bmq \nu}
    \tilde{B}_{\bmq \nu} \tilde{\mathcal{G}}^{\nu}_{\vck,\vckk} (\bmqq,\bmqq',\bmq) 
    \Bigg].
\end{align}
The corresponding expression in the exciton basis is:
\begin{align}
\label{eqn:exciton_etot}
    E_{\mathrm{xp}} =& \frac{1}{N_p}
    \sum_{\substack{ss'\\ \bmqq \bmqq'}} 
    A_{s\bmqq}^* A_{s'\bmqq'}
    \Bigg[ 
    E^0_{s\bmqq}
    \delta_{ss'}
    \delta_{\bm{Q} \bm{Q}'} 
    \nonumber \\
    &-\frac{1}{N_p}\sum_{\nu}
    B_{\bmqq-\bmqq' \nu} 
    \mathcal{G}_{ss'\nu}(\bmqq',\bmqq-\bmqq')
    \Bigg]
    \nonumber \\
    =&
    \varepsilon 
    + \frac{1}{N_p} \sum_{\bmq \nu} \abs{B_{\bmq\nu}}^2 \hbar \omega_{\bmq\nu}.
\end{align}
Furthermore, by combing Eqs.~(\ref{eqn:minimize_tau}), (\ref{eqn:expansion_transition})-(\ref{eqn:def_eph}), and (\ref{eqn:basis_transform_rev}), we can write the atomic displacements accompanying the excitonic polaron as:
\begin{align}
    \label{eqn:displacement}
    \Delta \tau_{\kpa}
    =
    -\frac{2}{N_p} \sum_{\bmq \nu}
    B_{\bmq \nu} \left(\frac{\hbar}{2M_{\kappa}\omega_{\bmq \nu}}\right)^{1/2}
    e_{\kappa \alpha, \nu}(\bmq) e^{i\bmq\cdot\bmrr_p}.
\end{align}
Based on this expression, the coefficients $B_{\bmq \nu}$ can be interpreted as the contributions of each normal vibrational mode to the excitonic polaron. This expression is also found in the case of the electron and hole polarons~\cite{sio2019ab}.

\section{Model Systems}
\label{sec:model}

To gain insight into the nature of the solutions of Eqs.~(\ref{eqn_rs:polaron_eqn_final})-(\ref{eqn_rs:g_transition}) and Eqs.~(\ref{eqn:explrneqn})-(\ref{eqn:exphg}), we start by considering the Wannier exciton model~\cite{wannier1937structure, mahan2000many} in the presence of Fr\"{o}hlich electron-phonon interactions~\cite{frohlich1950xx} or Holstein electron-phonon interactions~\cite{holstein1959studies}. This analysis will allow us to identify qualitative trends and to rationalize the \textit{ab initio} calculations presented in Sec.~\ref{sec:abinitio}.

\subsection{Wannier exciton model}
\label{sec:wannier}

In the Wannier model for excitons~\cite{mahan2000many}, the electronic structure is composed of one valence band and one conduction band, both of which are described in the effective mass approximation, as shown in Fig.~\ref{fig:illusstrate_wannier}(a). The electron and hole interact via an effective kernel that is given by the Coulomb interaction screened by the macroscopic dielectric constant $\epsilon^{\infty}$. With these choices, the effective Hamiltonian of the Wannier exciton reads~\cite{cardona2005fundamentals,fuchs2008efficient}:
\begin{align}
    \label{eqn_wf:bse_wf}
    &(H^0_{\mathrm{BSE}})_{v\bmk c\bmk+\bm{Q},v\bmk' c\bmk'+\bm{Q}} 
    \nonumber \\
    =&
    \left(
    \frac{\hbar^2 \abs{\bmk+\bmqq}^2}{2m_e} + \frac{\hbar^2 \abs{\bmk}^2}{2m_h} + E_g
    \right)
    \delta_{\bmk \bmk'}
    \nonumber \\
    &
    -
    \frac{e^2}{\epsilon_0\epsilon^{\infty}}\frac{1}{N_p \Omega} \frac{1}{\abs{\bmk'-\bmk}^2}.
\end{align}
In this expression, $\Omega$ is the volume of the primitive cell, $e$ is the electron charge, $\epsilon_0$ is the permittivity of vacuum, $m_e$ and $m_h$ are the electron and hole effective mass, respectively, and $E_g$ is the fundamental gap of the system.

Equation~(\ref{eqn_wf:bse_wf}) admits exact solutions~\cite{cardona2005fundamentals}. To see this, we transform the exciton eigenvector $a_{\bmk}^{\bmqq}$  into the Wannier representation: 
\begin{align}
    \label{eqn:a_to_w_wannier}
    \Phi(\bmrr_e,\bmrr_h)
    &=
    \frac{1}{N_p}
    \sum_{\bmk\bmqq} e^{i(\bmk+\bmqq) \cdot \bmrr_e}  e^{-i\bmk \cdot \bmrr_h} a_{\bmk}^{\bmqq},
\end{align}
having omitted the subscripts $vc$ and $s$ because we only have one valence band and one conduction band. With this transformation, the exciton wave function $\Omega(\bmr_e, \bmr_h)$ can be expressed as a linear combination of electron and hole Wannier functions, $w_{c\bmrr_e}
(\bmr_e)$ and $w^*_{v\bmrr_h}(\bmr_h)$:
\begin{align}
    \Omega(\bmr_e, \bmr_h)
    &= \frac{1}{\sqrt{N_p}}\sum_{\bmk\bmqq} a_{\bmk}^{\bmqq} 
    \psi_{c\bmk+\bmqq} (\bmr_e) \psi^*_{v\bmk}(\bmr_h)
    \nonumber \\
    &=\frac{1}{\sqrt{N_p}}
    \sum_{\bmrr_e \bmrr_h} \Phi(\bmrr_e,\bmrr_h)
    w_{c\bmrr_e}(\bmr_e) w^*_{v\bmrr_h}(\bmr_h).
\end{align}
We can now introduce the center-of-mass coordinate $\bmrr$ and the relative coordinate $\bmr$:
\begin{align}
    \bmrr = \frac{m_e \bmrr_e + m_h \bmrr_h}{m_e + m_h},~~
    \bmr = \bmrr_e - \bmrr_h,
\end{align}
as well as the total mass $M$ and the reduced mass $\mu$:
\begin{align}
    M = m_e + m_h,~~
    \frac{1}{\mu} = \frac{1}{m_e} + \frac{1}{m_h},
\end{align}
so as to recast the eigenvalue problem of the Wannier exciton into an hydrogenic Schr\"odinger equation~\cite{cardona2005fundamentals} [Fig.~\ref{fig:illusstrate_wannier}(b)]. The ground-state solution of this equation is the hydrogenic $1s$ wavefunction:
\begin{align}
    \Phi_{1s}(\bmrr_e, \bmrr_h) = \sqrt{\frac{1}{\pi a_0^3}}
    e^{i\bmqq \cdot \bmrr}
    e^{-{\abs{\bmr}}/{a_0}},
    \\
    E_{1s} = E_g + \frac{\hbar^2 \abs{\bmqq}^2}{2 M} 
    - \frac{\mu}{2} 
    \left(\frac{e^2}{4\pi\epsilon^\infty \epsilon_0 \hbar}\right)^2,
\end{align}
where $a_0 = {4\pi \epsilon^{\infty} \hbar^2} / {\mu e^2}$ is the exciton Bohr radius. Going back to the Bloch representation, we obtain the BSE coefficients for this wavefunction:
\begin{align}
    \label{eqn:1s_state}
    a_{\bmk}^{\bmqq}
    &=
    8\sqrt{\frac{\pi a_0^3}{\Omega}}
    \frac{1}{ \left( a_0^2\abs{\bmk + m_h\bmqq/M}^2 
    + 1 \right)^2}.
\end{align}

\begin{figure}
    \centering
    \includegraphics[width=0.48\textwidth]{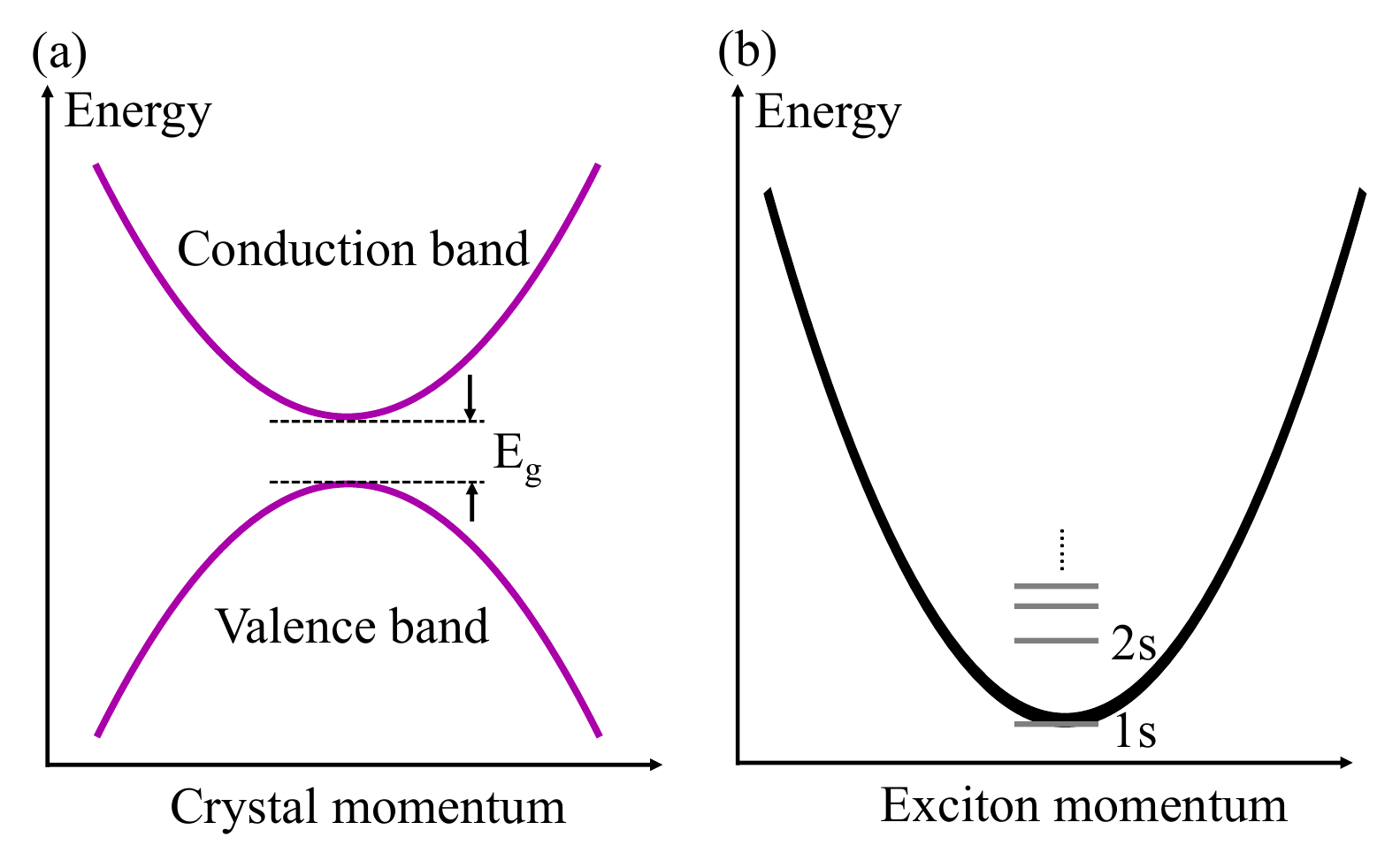}
    \caption{Schematic illustration of the Wannier model for excitons. 
    (a) In the Wannier model, there is one parabolic valence band and one parabolic conduction band, separated by the quasi-particle band gap $E_g$. Both bands are described by the effective mass approximation, and there is attractive Coulomb interaction between electrons and holes screened by the electronic dielectric constant $\epsilon^{\infty}.$
    (b) The Wannier model for excitons is analytically solvable. Upon changing the electron and hole coordinates into the center-of-mass reference frame, the components of the eigenstates in the relative coordinate are hydrogenic wavefunctions, whose energies are indicated by grey horizontal lines. The components of the eigenstates in the center-of-mass coordinate are planewaves, whose energies are indicated by the black parabola (for the $1s$ exciton). Accordingly, the free excitons are fully delocalized in the center-of-mass coordinate, and localized in the relative coordinate.}
    \label{fig:illusstrate_wannier}
\end{figure}

\subsection{Fr\"ohlich and Holstein electron-phonon couplings}
\label{sec:froh_hols_eph}

The Fr\"{o}hlich interaction is a widely-used analytical model to describe electrons coupled to long-wavelength longitudinal optical modes in polar materials~\cite{frohlich1950xx}. The Fr\"{o}hlich electron-phonon matrix element is given by~\cite{sio2019ab,lafuente2022ab}: 
\begin{align}
    \label{eqn_wf:frohlich}
    g(\bmq)=\frac{i}{\abs{\bmq}} 
    \sqrt{
    \frac{e^2}{4\pi\epsilon_0}
    \frac{4\pi}{\Omega}
    \frac{\hbar \omega_{\mathrm{LO}}}{2}
    \frac{1}{\kappa}
    },
\end{align}
where $\kappa=(1/\epsilon^\infty-1/\epsilon^0)^{-1}$~\footnote{The $\kappa$ appearing in the definition of the {Fr\"{o}hlich} electron-phonon matrix element should not be confused with the $\kappa$ used to deonote the atoms in $\Delta \tau_{\kpa}$.}, $\epsilon^0$ is the static dielectric constant including the ionic contribution, and $\omega_{\mathrm{LO}}$ is the frequency of the longitudinal optical mode. 
This model considers only intra-band couplings, and the coupling strength depends only on the phonon wavevector $\bmq$~\cite{verdi2015frohlich,sio2022unified}. Thus, in the case of the Wannier exciton model, the electron-phonon and hole-phonon coupling have exactly the same form as given by Eq.~(\ref{eqn_wf:frohlich}).

The Fr\"{o}hlich model is designed to describe long-range polar interactions, and does not capture short-range interactions~\cite{verdi2015frohlich}. To qualitatively analyze short-range electron-phonon couplings, we consider the Holstein model. In this model, the matrix elements are taken to be constant throughout the Brillouin zone~\cite{holstein1959studies}:
\begin{align}
    \label{eqn_wf:holstein}
    g_c(\bmq) = \frac{1}{\sqrt{\Omega}}g_c,~~g_v(\bmq) = \frac{1}{\sqrt{\Omega}}g_v,
\end{align}
with $g_c$ and $g_v$ being two materials-dependent parameters. In the following, we study how the Fr\"ohlich interaction and the Holstein interaction as defined by Eqs.~\eqref{eqn_wf:frohlich} and \eqref{eqn_wf:holstein} influence the formation of excitonic polarons.

\vspace*{10pt}

\subsection{Total energy in the transition basis approach}
\label{sec:etot_transition}

The transition basis approach outlines in Sec.~\ref{sec:transition} is particularly useful in the case of the Fr\"ohlich interaction to gain insight into the mechanisms that lead to the formation of the excitonic polaron. To see this, we evaluate the total energy of the excitonic polaron in by combining Eq.~(\ref{eqn:expansion_transition}) with Eq.~(\ref{eqn:a_to_w_wannier}):
\begin{widetext}
\begin{align}
    \label{eqn_wf:etot_density}
    E_{\mathrm{xp}} = 
    &-\frac{1}{N_p}\sum_{\bmrr_e,\bmrr_h}
    \Phi^*(\bmrr_e,\bmrr_h)
    \frac{\hbar^2\nabla^2_{\bmrr_e}}{2m_e} \Phi(\bmrr_e,\bmrr_h)
    - 
    \frac{e^2}{8\pi\epsilon_0} \frac{1}{\kappa} 
    \sum_{\bmrr_e,\bmrr_e'}
    \frac{n_e(\bmrr_e)n_e(\bmrr_e')}{\abs{\bmrr_e - \bmrr_e'}}
    \nonumber \\
    &-\frac{1}{N_p}\sum_{\bmrr_e,\bmrr_h}
    \Phi^*(\bmrr_e,\bmrr_h)
    \frac{\hbar^2\nabla^2_{\bmrr_h}}{2m_h} \Phi(\bmrr_e,\bmrr_h)
    - 
    \frac{e^2}{8\pi\epsilon_0} \frac{1}{\kappa} 
    \sum_{\bmrr_h,\bmrr_h'}
    \frac{n_h(\bmrr_h)n_h(\bmrr_h')}{\abs{\bmrr_h - \bmrr_h'}}
    \nonumber \\
    &
    -\frac{1}{N_p} \sum_{\bmrr_e,\bmrr_h}
    \frac{e^2}{4\pi \epsilon_0 \epsilon^{\infty}} 
    \frac{\abs{\Phi(\bmrr_e,\bmrr_h)}^2}{\abs{\bmrr_e-\bmrr_h}}
    +
    2\frac{e^2}{8\pi\epsilon_0} \frac{1}{\kappa} 
    \Bigg[
    \sum_{\bmrr_e,\bmrr_h}
    \frac{n_h(\bmrr_h)n_e(\bmrr_e)}{\abs{\bmrr_h - \bmrr_e}}
    \Bigg]
    + E_g,
\end{align}  
\end{widetext}
having defined the electron and hole densities as:
\begin{align}
    \label{eqn_wf:charge_density}
    n_e(\bmrr_e) &= \frac{1}{N_p} \sum_{\bmrr_h} \abs{\Phi(\bmrr_e,\bmrr_h)}^2,
\end{align}
\begin{align}
    n_h(\bmrr_h) &= \frac{1}{N_p} \sum_{\bmrr_e} \abs{\Phi(\bmrr_e,\bmrr_h)}^2.
\end{align}
In Eq.~(\ref{eqn_wf:etot_density}), terms are arranged in a such a way that direct comparison with the Landau-Pekar model of polarons can be made~\cite{landau1933electron, pekar1946local, sio2019ab}: The first line describes the energetics of an electron polaron, and the second line describes the energetics of a hole polaron. The first term on the third line, which contains purely electronic screening, describes the Coulomb attraction between the electron and hole, which binds the electron and hole polarons together. The second term on the third line, which contains the ionic contribution to the screening, describes the weakening of the electron and hole interactions with the lattice resulting from charge compensation in the exciton state. Thus, the formation of excitonic polaron can be thought of as a two-step process, in which the first step is the formation of independent electron polaron and hole polaron, and the second step involves the binding of these polarons by their mutual Coulomb attraction, accompanied by the weakening of the lattice distortion from the partial cancellation of the electron and hole charge densities.

An alternative way to interpret Eq.~(\ref{eqn_wf:etot_density}) is obtained by combining together all terms proportional to $1/\kappa$:
\begin{widetext}
\begin{align}
    \label{eqn_wf:etot_density2}
    E_{\mathrm{xp}} = 
    &-\frac{1}{N_p}\sum_{\bmrr_e,\bmrr_h}
    \Phi^*(\bmrr_e,\bmrr_h)
    \left[\frac{\hbar^2\nabla^2_{\bmrr_e}}{2m_e} + \frac{\hbar^2\nabla^2_{\bmrr_h}}{2m_h}\right]\Phi(\bmrr_e,\bmrr_h)
    -\frac{1}{N_p} \sum_{\bmrr_e,\bmrr_h}
    \frac{e^2}{4\pi \epsilon_0 \epsilon^{\infty}} 
    \frac{\abs{\Phi(\bmrr_e,\bmrr_h)}^2}{\abs{\bmrr_e-\bmrr_h}}
    \nonumber \\
    &
    - \frac{e^2}{8\pi\epsilon_0} \frac{1}{\kappa} 
    \sum_{\bmrr,\bmrr'}
    \frac{\Delta n(\bmrr) \Delta n(\bmrr')}{\abs{\bmrr - \bmrr'}}
    + E_g,
\end{align}  
\end{widetext}
where $\Delta n(\bmrr) = n_e(\bmrr) - n_h(\bmrr)$ is the net charge density of the exciton. 
{Note that this charge density can be positive or negative at different regions, leading to partial cancellation in the last term of Eq.~(\ref{eqn_wf:etot_density2}) and a weakening of the polaronic stabilization mechanism. 
This indicates that it might be harder to form excitonic polarons than charged polarons.}
In this form, the first line describes the standard Wannier exciton energy, in the absence electron-phonon interactions. The second line describes the stabilization energy provided by the interaction of the net charge density of the exciton with the ionic lattice, as in the Landau-Pekar model of polarons.
{We note that in both Eq.~(\ref{eqn_wf:etot_density}) and Eq.~(\ref{eqn_wf:etot_density2}), the total energy does not depend on the phonon frequency $\omega_\mathrm{LO}$ in the Fr\"{o}hlich electron-phonon coupling [Eq.~(\ref{eqn_wf:frohlich})]. 
This is because the $\omega_\mathrm{LO}$ is canceled out by the prefactor in $B_{\bmqq\nu}$ [Eq.~(\ref{eqn_rs:g_transition})].
The absence of the phonon frequency in the energy is a consequence of the adiabatic approximation, and is similar to what is found in the strong-coupling solution of the Fr\"{o}hlich polaron problem~\cite{sio2019ab, devreese2010frohlich}.}

Overall, Eqs.~\eqref{eqn_wf:etot_density} and \eqref{eqn_wf:etot_density2} show that we can think of excitonic polarons in the Wannier-Fr\"ohlich model in one of two ways: (i) a particle formed from the binding of electron and hole Pekar polarons; (ii) a particle formed from the interaction of a Wannier exciton with the ionic lattice.

\subsection{Total energy in the exciton basis approach}
\label{sec:etot_exciton}

\subsubsection{Fr\"ohlich electron-phonon interactions}\label{subsub.fr}

While Eqs.~\eqref{eqn_wf:etot_density} and \eqref{eqn_wf:etot_density2} are useful to conceptualize the formation of excitonic polarons in the Wannier-Fr\"ohlich model, they are not a good starting point to find an exact solution to the problem. In fact, unlike in the Landau-Pekar model, where a hydrogenic variational ansatz yields quantitatively accurate solutions for the ground state~\cite{sio2019ab}, identifying a useful variational ansatz for $\Phi(\bmrr_e, \bmrr_h)$ to be used in Eq.~(\ref{eqn_wf:etot_density}) is challenging.

To overcome this difficulty, we move to the exciton basis representation of Sec.~\ref{sec:exciton} and include a single $1s$ exciton band in Eq.~\eqref{eq.exbasis}. For the case of Fr\"{o}hlich interactions, we substitute Eqs.~(\ref{eqn:1s_state})-(\ref{eqn_wf:frohlich}) into Eq.~(\ref{eqn:exphg}), to find the Fr\"{o}hlich exciton-phonon coupling matrix element:
\begin{align}
    \label{eqn:exphg_froe}
    \mathcal{G}^{\mathrm{F}}(\bmqq, \bmq) = 
    g(\bmq)
    \left[\frac{1}{ \left(a_0^2 b^2|\bmq|^2/4 + 1 \right)^2}
    -
    \frac{1}{ \left(a_0^2 a^2 |\bmq|^2/4 + 1 \right)^2}
    \right],
\end{align}
where $g(\bmq)$ is the Fr\"{o}hlich electron-phonon coupling matrix element from Eq.~\eqref{eqn_wf:frohlich}, $a = m_e/(m_e + m_h)$, and $b = m_h/(m_e + m_h)$.

From Eq.~(\ref{eqn:exphg_froe}) and Fig.~\ref{fig:analytical_exphg} one can see that, unlike the standard Fr\"{o}hlich \textit{electron}-phonon interaction which diverges as $\bmq\rightarrow0$, the Fr\"{o}hlich \textit{exciton}-phonon interaction tends to vanish in the long wavelength limit.

In order to evaluate the total energy according to Eqs.~(\ref{eqn:bmat}) and (\ref{eqn:exciton_etot}), we need an ansatz for the coefficients $A_{s\bmqq}$ of the excitonic polaron. We consider the following hydrogenic ansatz:
\begin{align}
    \label{eqn:explrn_ansstz}
    A_{\bmqq} 
    &= 8\sqrt{\frac{\pi r_p^3}{\Omega}} \frac{1}{(r_p^2|\bmqq|^2 + 1)^2}.
\end{align}    
With this choice, a large $r_p$ indicates that the excitonic polaron is mostly formed by excitons near the zone center, and {\it vice versa}. In the extreme case $r_p\rightarrow\infty$, the exciton polaron is completely delocalized and reduces to a $\Gamma$-point Wannier exciton.

Using Eq.~(\ref{eqn:explrn_ansstz}), we find the following analytical expression for the total energy of the Wannier-Fr\"ohlich excitonic polaron:
\begin{align}
    E_{\mathrm{xp}} &= E_{\rm el} + E^{\mathrm{F}}_{\rm ph},
\end{align}
where
\begin{align}
    E_{\rm el}
    &=
    \frac{\hbar^2 }{2(m_e + m_h) r^2_p},
    \label{eqn:electronic_model} \\
    E^{\mathrm{F}}_{\rm ph}
    &=
    -\frac{1}{4}
    \frac{1}{\pi^2}
    \frac{e^2}{\epsilon_0 \kappa}
    \frac{a_0^4 (a-b)^2 \pi}{16(a a_0 + r_p)^7 (b a_0 + r_p)^7}
    \sum_{i=1}^{10} t_i. \label{eqn:ene_ph_froe}
\end{align}
The explicit expressions for the terms $t_i$ in the last equation are provided in App.~\ref{app:full_froe}. These terms only depend on the coefficients $a$, $b$, and on the variational parameter $r_p$. 
As expected, the kinetic energy contribution is positive for all values of $r_p$, and tends to favor delocalization; conversely, the phonon contribution is negative and favors localization. Using Eqs.~\eqref{eqn:electronic_model} and \eqref{eqn:ene_ph_froe}, we can find the radius $r_p$ that minimizes the total energy numerically.

\begin{figure}
    \centering
    \includegraphics[width=0.5\textwidth]{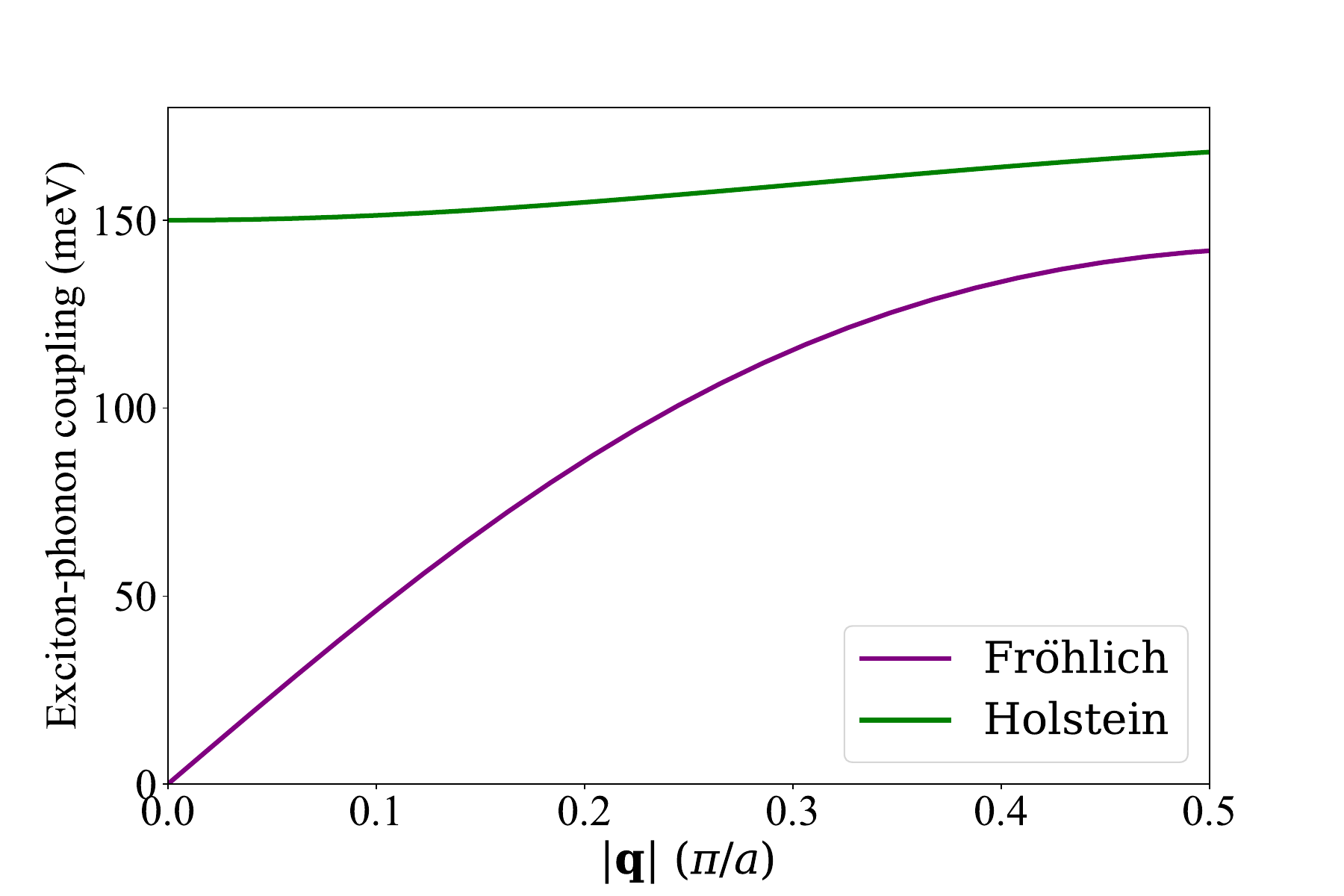}
    \caption{Fr\"{o}hlich and Holstein exciton-phonon couplings for Wannier excitons. For Fr\"{o}hlich type interactions (purple line), the exciton-phonon matrix element vanishes as $\bmq \rightarrow 0$. This trend is in contrast to the Fr\"{o}hlich electron-phonon interaction, which diverges in the same limit. In the case of the Holstein exciton-phonon interaction (green line), the matrix element shows little dispersion throughout the Brillouin zone, and remains finite in the limit $\bmq \rightarrow 0$. The parameters used for this plot are summarized in Tab.~\ref{tab:parameters}.
    }
    \label{fig:analytical_exphg}
\end{figure}

If we choose parameters that are typical for lithium fluoride, as summarized in Tab.~\ref{tab:parameters}~\cite{sio2019ab}, which correspond to the ratio $m_h=5m_e$, we find that the the localization energy $E_{\rm ph}$ is too small to overcome the delocalization energy $E_{\rm el}$. As a result, the total energy remains positive for all values of $r_p$, and reaches its minimum for the fully-delocalized solution [solid line in Fig.~\ref{fig:model_total_energy}(a)]. In this case there is no excitonic polaron.

Since $E^{\mathrm{F}}_{\rm ph}$ in Eq.~\eqref{eqn:ene_ph_froe} is controlled by $m_e$, $m_h$, and $\kappa$, in Fig.~\ref{fig:model_total_energy}(a) we also explore what happens when we artificially increase the hole effective mass to $m_h=15m_e$ (dashed line). In this case, we find a local minimum in the total energy landscape indicating the formation of an excitonic polaron. This observation indicates that, in the case of Fr\"{o}hlich interactions, there exists a critical condition that $m_e$, $m_h$, and $\kappa$ must fulfil for excitonic polarons to form. Intuitively, this is understood on the grounds that the driving force for localization is the electrostatic interaction between the net exciton charge and the ionic lattice [cf. Eqs.~\eqref{eqn_wf:etot_density} and \eqref{eqn_wf:etot_density2}], therefore localization requires a large difference in effective masses, or a large ionic contribution to the dielectric screening, or both. 

We note that a difference in electron and hole masses as large as that considered in Fig.~\ref{fig:model_total_energy}(a) is unrealistic for most materials. However, we expect that in the presence of multiple longitudinal optical phonons and multiple bands, the contributions from these different channels should add up to enable the formation of excitonic polarons even when the difference between electron and hole masses is not as pronounced.

\subsubsection{Holstein electron-phonon interactions}

In the case of the Holstein electron-phonon interaction [Eq.~\eqref{eqn_wf:holstein}], the exciton-phonon matrix elements take the following form:
\begin{align}
    \label{eqn:exphg_hols}
    \mathcal{G}^{\mathrm{H}}(\bmqq, \bmq) = 
    \frac{1}{\sqrt{\Omega}}
    \left[\frac{g_c}{ \left(a_0^2 b^2|\bmq|^2/4 + 1 \right)^2}
    -
    \frac{g_v}{ \left(a_0^2 a^2|\bmq|^2/4 + 1 \right)^2}
    \right].
\end{align}
A plot of these matrix elements for parameters corresponding to LiF is shown in Fig.~\ref{fig:analytical_exphg}.

By using the same hydrogenic wavefunction ansatz as in Eq.~(\ref{eqn:explrn_ansstz}), we obatain the phonon contribution to the total energy of the excitonic polaron as:
\begin{align}
    \label{eqn:ene_ph_hols}
    E^{\mathrm{H}}_{\rm ph} 
    &=
    -\frac{1}{2 \hbar \omega_{\mathrm{LO}}}
    \frac{1}{\pi^2} 
    \int_0^{\infty}{dQ} 
    \frac{Q^2}{ \left(r_p^2 Q^2/4 + 1 \right)^4}
    \nonumber \\
    &\times
    \left[\frac{g_c}{ \left(a_0^2 b^2 Q^2/4 + 1 \right)^2}
    -
    \frac{g_v}{ \left(a_0^2 a^2 Q^2/4 + 1 \right)^2}
    \right]^2.
\end{align}
Evaluating this integral is cumbersome but possible, we provide the explicit expression in App.~\ref{app:full_hols}. As in the case of Fr\"ohlich interactions, it is possible to determine numerically the radius $r_p$ that minimized the total energy for a give choice of the parameters $m_e$, $m_h$, $g_v$, $g_c$, and $\omega_{\rm LO}$.

Using the same effective masses as for LiF in the previous section, we have found that typical values of coupling constants, such as $g_c/\sqrt{\Omega}=50$~meV and $g_v/\sqrt{\Omega}=200$~meV do not lead to the formation of excitonic polarons. However, as we show in Fig.~\ref{fig:model_total_energy}(b), in the presence of both Fr\"{o}hlich and Holstein interactions, the total energy landscape exhibits a local minimum, signaling the formation of excitonic polarons. This finding indicates that, for materials where both the long-range and short-range electron-phonon interactions play a role, the formation of excitonic polarons is possible under realistic parameters for the electron-phonon couplings matrix elements, band effective masses, and dielectric constants.

\begin{table}[]
\begin{tabular}{ *{8}{>{\centering\arraybackslash}m{0.05\textwidth}} }
\toprule
\toprule
$m_e$ & $m_h$ & $\epsilon^{\infty}$ & $\epsilon^{0}$ & $\Omega$  & $g_c/\sqrt{\Omega}$ & $g_v/\sqrt{\Omega}$ & $\hbar \omega_{\mathrm{LO}}$\\ 
\midrule
0.88 & 4.4 & 2.04 & 10.62 & 27 & 50 & 200 & 77 \\
\bottomrule
\bottomrule
\end{tabular}
\caption{Parameters used to solve the excitonic polaron equations for the Wannier exciton model with Fr\"{o}hlich or Holstein electron-phonon couplings. The effective masses are given in units of the bare electron mass $m_0$. The unit cell volume $\Omega$ is in $\mathrm{\AA}^{3}$. All other quantities are in meV.}
\label{tab:parameters}
\end{table}

\begin{figure*}
    \centering
    \includegraphics[width=1.0\textwidth]{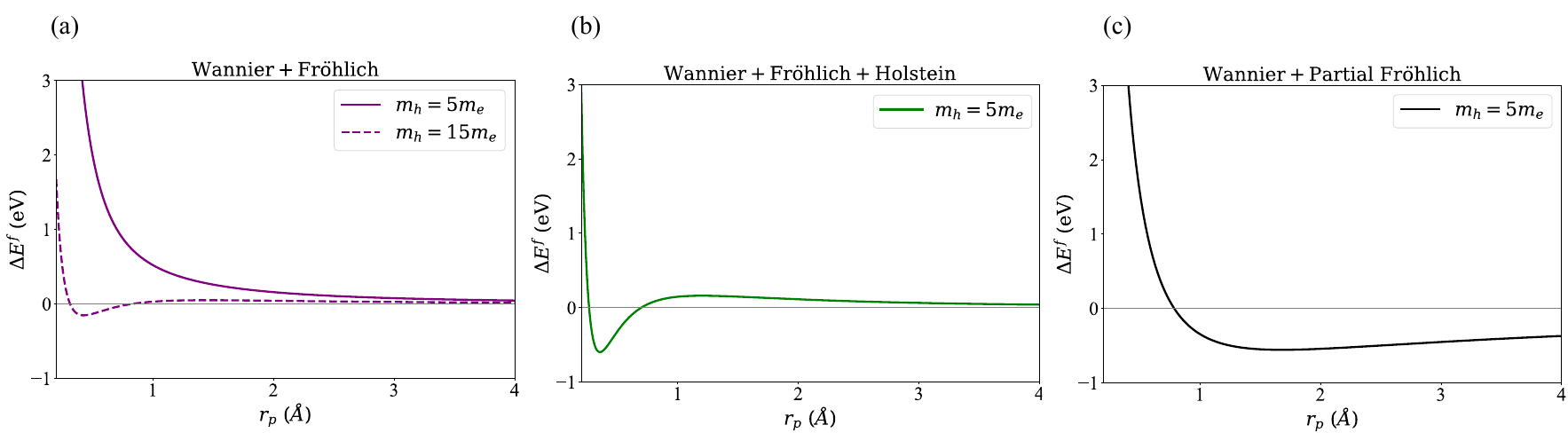}
    \caption{Formation energy of the excitonic polaron relative to the total energy of the free exciton, for various model systems. (a) Wannier exciton with Fr\"{o}hlich electron-phonon interaction. When $m_h = 5 m_e$, the minimum energy solution corresponds to a fully-delocalized exciton (solid line). When the hole mass is much larger ($m_h = 15 m_e$), the minimum energy corresponds to a localized solution (dashed line). This observation indicates that a large difference in effective masses between electrons and holes favors the formation of the excitonic polaron. (b) Wannier exciton with both Fr\"{o}hlich electron-phonon interaction and Holstein electron-phonon interaction. With the help of short-range Holstein-type interactions, localized solutions emerge when there is a significant difference in electron-phonon coupling and hole-phonon coupling (in this example, 50~meV and 200~meV, respectively). (c) Wannier exciton with artificially-enhanced Fr\"{o}hlich exciton-phonon interaction. In this example, the electron-phonon coupling matrix element $g_{cc'\nu}(\bmk+\bmqq,\bmq)$ in Eq.~(\ref{eqn:exphg}) is manually set to zero. This observation provides us with a strategy to initialize the iterative minimization of the \textit{ab initio} excitonic polaron equations.}
    \label{fig:model_total_energy}
\end{figure*}

\subsection{Implications for first-principles calculations}
\label{sec:implication}

Close inspection of Figs.~\ref{fig:model_total_energy}(a) and ~\ref{fig:model_total_energy}(b) reveals that, when an excitonic polaron solution exists, there is also a barrier between this localized solution and the fully-delocalized exciton solution. Such a barrier may pose a challenge to first-principles calculations, because if the initial guess for the self-consistent calculation is far away from the localized solution, then the minimization procedure is likely to fall back to the fully delocalized solution. Thus, devising an effective strategy to identify potential localized solutions is crucial when performing first-principles calculations.

To address this challenge, we go back to the definition of the exciton-phonon coupling matrix element Eq.~(\ref{eqn:exphg}). There, we see that first two terms on the right-hand-side appear with opposite signs, hence they tend to cancel each other. This partial cancellation reflects the fact that it is the net charge of the exciton that generates forces on the ions. This cancellation also echoes the difficulty of finding localized solutions in the Fr\"{o}hlich model, as discussed in Sec.~\ref{subsub.fr}.

Based on the above observation, it should be easier to find localized solutions of Eqs.~(\ref{eqn:explrneqn})-(\ref{eqn:exphg}) if we artificially set to zero the first term in Eq.~(\ref{eqn:exphg}). Figure~\ref{fig:model_total_energy}(c) shows that, indeed, with this modification of the matrix elements, the potential energy landscape exhibits a single minimum at the localized solution in the case of the Wannier-Fr\"ohlich model. While this alteration of the matrix elements has no physical meaning {\it per se}, it offers a simple and useful strategy to initialize the self-consistent solution of Eqs.~(\ref{eqn:explrneqn}) in \textit{ab initio} calculations. In practice, we can solve Eqs.~(\ref{eqn:explrneqn}) in two steps: (i) first, we obtain an artificially-localized solution by setting electron-phonon interaction or hole-phonon interaction to zero; (ii) second, we use this artificially-localized solution as a seed for a second run with the correct matrix elements including both electron-phonon and hole-phonon interactions. In Sec.~\ref{sec:abinitio} we show how this two-step approach allows us to find excitonic polaron solutions in first-principles calculations of real materials.

\section{First-principles calculations}
\label{sec:abinitio}

To implement the methodology described in Sec.~\ref{sec:theory}, we choose to proceed with Eqs.~(\ref{eqn:explrneqn})-(\ref{eqn:exphg}) since they show a more favorable scaling with the number of bands and grid points in the Brillouin zone. In this section, we first discuss the gauge invariance of our formalism and the hermicity of the effective excitonic polaron Hamiltonian. Second, we briefly touch upon calculations of finite-momentum excitons. Third, we discuss our results for lithium fluoride (LiF). An additional application of this methodology is presented in the companion manuscript~\cite{dai2023explrn_prl}, for the vacancy-ordered double perovskite \ch{Cs2ZrBr6}.

\subsection{Gauge invariance and hermiticity}
\label{sec:gauge}

The present formalism carries gauge freedoms in several points, namely the Kohn-Sham wave functions, the phonon eigenvectors, and the BSE eigenvectors, which are all obtained by matrix diagonalization. In the absence of degeneracy, all these quantities are defined modulo a complex phase; in the presence of degeneracy, any unitary transformation within the degenerate subspace is admissible. As a physical observable, the total energy of the excitonic polaron and its associated eigenvalue must be invariant with respect to these freedoms. Thus, we need to make sure that our formalism is indeed gauge-invariant.

We first analyze the case of a system without any degeneracy. We consider a change of phase of the Kohn-Sham wavefunctions: $\psi_{n\bmk}(\bmr)\rightarrow e^{i\phi_n(\bmk)}\psi_{n\bmk}$. With this change, the BSE eigenvectors $a_{\vck}^{s\bmqq}$ acquire a corresponding complex phase:
\begin{align}
    a_{\vck}^{s\bmqq}
    \rightarrow
    e^{i\phi_v(\bmk)} e^{-i\phi_c(\bmk+\bmqq)}
    a_{\vck}^{s\bmqq}.
\end{align}
Similarly, the electron-phonon matrix elements also acquire a complex phase:
\begin{align}
    &g_{mn\nu}(\bmk,\bmq)
    \rightarrow
    e^{-i\phi_{m}(\bmk+\bmq)} e^{i\phi_{n}(\bmk)}
    g_{mn\nu}(\bmk,\bmq).
\end{align}
These phases cancel out when the BSE eigenvectors and the electron-phonon matrix elements are used in Eq.~(\ref{eqn:exphg}), therefore the exciton-phonon coupling matrix elements are independent of the phase of Kohn-Sham states.

A similar reasoning applies to the vibrational eigenmodes. Upon introducing a complex phase via $e_{\kappa \alpha, \nu}(\bmq) \rightarrow e^{i\phi_{\nu}(\bmq)}e_{\kappa \alpha, \nu}(\bmq)$, the electron-phonon matrix elements and therefore the exciton-phonon matrix elements acquire a corresponding phase:
\begin{align}
    g_{mn\nu}(\bmk,\bmq)
    &\rightarrow
    e^{i\phi_{\nu}(\bmq)}
    g_{mn\nu}(\bmk,\bmq),
    \nonumber \\
    \mathcal{G}_{ss'\nu}(\bmqq,\bmq)
    &\rightarrow
    e^{i\phi_{\nu}(\bmq)}
    \mathcal{G}_{ss'\nu}(\bmqq,\bmq).
\end{align}
At the same time, the coefficients $B_{\bmqq\nu}$ in Eq.~(\ref{eqn:bmat}) also acquire a phase through the exciton-phonon couplings $\mathcal{G}^*_{ss'\nu}(\bmqq,\bmq)$:
\begin{align}
    B_{\bmqq\nu}
    \rightarrow
    e^{-i\phi_{\nu}(\bmqq)}B_{\bmqq\nu}.
\end{align}
Thus, the phase factor $\phi_\nu(\bmq)$ cancels out in Eq.~(\ref{eqn:explrneqn}), making the present formalism independent of the phases of vibrational eigenmodes.

In the case of the BSE eigenvectors, the change of phase $a_{\vck}^{s\bmqq}\rightarrow e^{i\phi_s(\bmqq)}a_{\vck}^{s\bmqq}$ leads to the following modification of the exciton-phonon matrix elements:
\begin{align}
    \mathcal{G}_{ss'\nu}(\bmqq,\bmq)
    &\rightarrow
    e^{-i\phi_{s}(\bmqq+\bmq)}e^{i\phi_{s'}(\bmqq)}
    \mathcal{G}_{ss'\nu}(\bmqq,\bmq).
\end{align}
As a result, the the coefficients $B_{\bmqq\nu}$ remain unchanged:
\begin{align}
    \label{eqn:bmat_gauge}
    B_{\bmqq \nu}  
    \rightarrow&
    \frac{1}{N_p \hbar \omega_{\bmqq \nu}} 
    \sum_{\substack{ss' \bmqq'}}
    e^{i\phi_{s'}(\bmqq')} e^{-i\phi_{s}(\bmqq'+\bmqq)}
    A_{s'\bmqq'}^* A_{s\bmqq'+\bmqq}
    \nonumber \\
    &\times 
    e^{i\phi_{s}(\bmqq+\bmqq')} e^{-i\phi_{s'}(\bmqq')} 
    \mathcal{G}^*_{ss'\nu}( \bmqq', \bmqq)
    \nonumber \\
    =&
    B_{\bmqq \nu},
\end{align}
and from Eq.~(\ref{eqn:explrneqn}) we see that the eigenvector of the transformed excitonic polaron Hamiltonian becomes $e^{-i\phi_s(\bmqq)}A_{s\bmqq}$:
\begin{eqnarray}
    &&\sum_{s'\bmqq'} 
    \bigg[ 
    E^0_{s\bmqq}
    \delta_{ss'}
    \delta_{\bm{Q} \bm{Q}'} 
    -\frac{2}{N_p}\sum_{\nu}
    B_{\bmqq-\bmqq' \nu} \mathcal{G}_{ss'\nu}(\bmqq',\bmqq-\bmqq')
    \bigg] 
    \nonumber \\ && \hspace{20pt}\times A_{s'\bmqq'} e^{-i\phi_s(\bmqq)}
    = \varepsilon A_{s\bmqq} e^{-i\phi_s(\bmqq)}.
\end{eqnarray}
Clearly this change of phase of the solution coefficients does not alter the eigenvalue and the total energy of the excitonic polaron.

We now analyze how the formalism is affected by unitary transformations within degenerate subspaces.
We notice that, in Eq.~(\ref{eqn:exphg}), the indices of electron bands are always repeated twice. That is, whenever we have the band index $n$ in the ket $\ket{n\bmk}$, we also have a bra $\bra{n\bmk}$ in the same expression, and these indices are summed over. The summation over these indices yields the the resolution of identity within the degenerate subspace, $\sum_{j} \ket{n_j\bmk} \bra{n_j\bmk}=\hat{I}_n$, which is gauge invariant. The same reasoning can be carried out for degeneracies in the phonon eigenvalues and in the BSE eigenvalues. The result is that the eigenvalue and the total energy of the excitonic polaron are gauge-invariant under unitary transformations within degenerate subspaces of electrons, phonons, and excitons.

It remains to show that the excitonic polaron Hamiltonian is Hermitian. To this end, we swap the exciton band indices and exciton momenta:
\begin{align}\label{eq.tmp.0}
    &H_{s'\bmqq', s\bmqq}     
    \nonumber \\
    =&
    E^0_{s'\bmqq}
    \delta_{s's}
    \delta_{\bm{Q'} \bm{Q}} 
    -\frac{2}{N_p}\sum_{\nu}
    B_{\bmqq'-\bmqq \nu} \mathcal{G}_{s's\nu}(\bmqq,\bmqq'-\bmqq).
\end{align}
From the definition of exciton-phonon coupling matrix element in Eq.~(\ref{eqn:exphg}), we have:
\begin{align}\label{eq.tmp.1}
    &\mathcal{G}_{s's\nu}(\bmqq,\bmqq'-\bmqq)
    \nonumber \\
    =&
    \sum_{v\bmk} 
    \sum_{cc'}
    a_{\vck}^{s'\bmqq'*} 
    a_{vc'\bmk}^{s\bmqq}
    g_{cc'\nu} (\bmk+\bmqq,\bmqq' - \bmqq)
    \nonumber \\
    &-
    \sum_{c\bmk}
    \sum_{vv'}
    a_{\vck}^{s'\bmqq'*} 
    a_{v'c\bmk + \bmqq' - \bmqq}^{s\bmqq}
    g_{v'v\nu} (\bmk,\bmqq' - \bmqq).
\end{align}
If we exchange the summation indices $c$ and $c'$, as well as the indices $v$ and $v'$, define $\bmk' = \bmk+\bmqq'-\bmqq$, and change $\bmk'$ into $\bmk$ at the end, we obtain:
\begin{align}\label{eq.tmp.2}
    &\mathcal{G}_{s's\nu}(\bmqq,\bmqq'-\bmqq)
    \nonumber \\
    =&
    \sum_{\vck} a_{\vck}^{s\bmqq} 
    \bigg[
    \sum_{c'}
    g_{cc'\nu} (\bmk+\bmqq',\bmqq - \bmqq')
    a_{vc'\bmk}^{s'\bmqq'}
    \nonumber \\
    &-
    \sum_{v'}
    g_{v'v\nu} (\bmk,\bmqq - \bmqq') 
    a_{v'c\bmk + \bmqq - \bmqq'}^{s'\bmqq'}
    \bigg]^* e^{i\phi_\nu(\bmqq-\bmqq')} 
    \nonumber \\
    = \,&
    \mathcal{G}^*_{ss'\nu}(\bmqq',\bmqq-\bmqq') e^{i\phi_\nu(\bmqq-\bmqq')},
\end{align}
where $\phi_\nu(\bmqq-\bmqq')$ is the relative phase between an eigenmode and its time-reversal pair, that is $ e_{\kappa\alpha,\nu}(-\bmq)= [e^{i\phi_\nu(\bmq)} e_{\kappa\alpha,\nu}(\bmq)]^*$. This phase factor depends on the phonon branch $\nu$ and phonon wavevector, but it cancels out when taking the product $ B_{\bmqq'-\bmqq \nu} \mathcal{G}_{s's\nu}(\bmqq,\bmqq'-\bmqq)$. In fact, if we carry out the same steps as in Eqs.~\eqref{eq.tmp.1}-\eqref{eq.tmp.2} for the coefficients $B_{\bmqq'-\bmqq \nu}$, we find:
\begin{align}
    &B_{\bmqq'-\bmqq \nu} \mathcal{G}_{s's\nu}(\bmqq,\bmqq'-\bmqq) 
    \nonumber \\
    =& 
    \Big[B_{\bmqq-\bmqq' \nu} \mathcal{G}_{ss'\nu}(\bmqq',\bmqq-\bmqq') \Big]^*.
\end{align}
In combination with Eq.~\eqref{eq.tmp.0}, this identity shows that the excitonic polaron Hamiltonian is Hermitian: $H_{s'\bmqq', s\bmqq} = H^*_{s\bmqq, s'\bmqq'}$.

We emphasize that, in the above proof, we do not require the relation $e_{\kappa \alpha, \nu}(-\bmq)=e^*_{\kappa \alpha, \nu}(\bmq)$, which is generally not satisfied in density functional perturbation theory (DFPT) because the calculations for $\bmq$ and $-\bmq$ are performed independently~\cite{baroni2001phonons}.

\subsection{Bethe-Salpeter equations with finite exciton momentum}

The solution of Eqs.~(\ref{eqn:explrneqn})-(\ref{eqn:exphg}) requires the knowledge of exciton states with finite momenta [Eq.~(\ref{eqn:basis_transform})]. These eigenstates are obtained by solving the BSE by including matrix elements with finite momentum transfer:
\begin{align}
    \int_{sc} d\bmr
    &\psi^*_{c\bmk+\bmqq}(\bmr_e) \psi_{v\bmk}(\bmr_h)
    K_{\mathrm{BSE}}^0(\fourr)
    \nonumber \\
    &\times 
    \psi_{c'\bmk'+\bmqq}(\bmr_e') \psi_{v'\bmk'}(\bmr_h').
\end{align}
Finite-momentum BSE calculations have become possible during the past decade~\cite{deslippe2012berkeleygw,sangalli2019many,cudazzo2016exciton}. For example, calculations of finite-momentum excitons for 2D materials such as \ch{MoS2}, graphane, and \ch{BN} have been reported~\cite{qiu2015nonanalyticity, cudazzo2016exciton}. Since each wavevector $\bmqq$ is computed independently, the computational cost scales linearly with the density of the uniform $\bmqq$-grid, or equivalently with the size of the BvK supercell.

From Eq.~(\ref{eqn:exphg}), it clear that the $\bmk$-grid, $\bmq$-grid, and $\bmqq$-grid need to be commensurate with each other. Since it is harder to converge BSE calculations as compared to  electronic and phonon calculations, it is advantageous to first converge the BSE calculations with respect to the Brillouin-zone $\bmk$-grid, and then to choose the size of the $\bmq$-grid and of the $\bmqq$-grid based on the expected size of the excitonic polarons of interest.

For the analysis of gauge invariance carried out in Sec.~\ref{sec:gauge} to hold, in practical calculations it is critical to ensure that the same set of electronic wave functions be employed in the evaluation of the BSE kernel and of the electron-phonon matrix elements. Were this not the case, the phase relations used in Sec.~\ref{sec:gauge} to prove gauge invariance would not be valid, and calculations results would be incorrect and unpredictable.

We also note that, in the implementation of Eqs.~(\ref{eqn:explrneqn})-(\ref{eqn:exphg}), it is important to pay attention to the conventions adopted by different \textit{ab initio} software packages. For example, in the \textsc{BerkeleyGW} code, the transition basis is defined as~\cite{deslippe2012berkeleygw}:
\begin{align}
    \label{eqn:basis_bgw}
    \tilde{\Psi}_{s\bmqq}(\bmr_e,\bmr_h)
    =
    \sum_{\vck}\tilde{a}_{\vck}^{s\bmqq}
    \psi^*_{v\bmk+\bmqq}(\bmr_h)
    \psi_{c\bmk}(\bmr_e).
\end{align}
On comparing with Eq.~(\ref{eqn:basis_transform}), we see that $\tilde{\Psi}_{s\bmqq}(\bmr_e,\bmr_h)$ actually represents an exciton state with momentum $-\bmqq$: $\tilde{\Psi}_{s\bmqq}(\bmr_e,\bmr_h) = \Psi_{s-\bmqq}(\bmr_e,\bmr_h)$. Thus, when implementing the present formalism in conjunction with the \textsc{BerkeleyGW} code, it is necessary to convert the exciton eigenvectors according to the following relation:
\begin{align}
    a_{\vck}^{s\bmqq}
    =\tilde{a}_{\vck+\bmqq}^{s-\bmqq}.
\end{align}

\subsection{Results: lithium fluoride}

In this section, we demonstrate the use of the present formalism in {\it ab initio} calculations of excitonic polarons in LiF. LiF is a simple cubic insulator that crystallizes in the rock-salt structure, and exhibits strong Fr\"ohlich electron-phonon couplings. LiF compound hosts small hole polarons and large electron polarons~\cite{sio2019ab}, and has been proposed to also host excitonic polarons~\cite{nakonechnyi2005low}. Previous work has shown that a relatively coarse $\bmk$-grid is sufficient to converge BSE calculations of excitons in LiF~\cite{onida2002electronic}, making it an ideal candidate for testing our formalism and studying the convergence behavior.

\subsubsection{Computational details}

To obtain the optimized structure, Kohn-Sham states and energies, and phonon eigenmodes and frequencies, we perform DFT and DFPT calculations using the \textsc{Quantum Espresso} package~\cite{giannozzi2009quantum, giannozzi2017advanced}. We employ the PBE generalized-gradient approximation to the exchange and correlation functional~\cite{perdew1996generalized}, norm-conserving pseudopotentials~\cite{hamann2013optimized,van2018pseudodojo}, and a planewaves kinetic energy cutoff of 100~Ry. The convergence threshold for self-consistent calculations is $10^{-12}${Ry}, and for DFPT the convergence threshold is $10^{-14}\mathrm{(Ry)}^2$.
We use \textsc{EPW}~\cite{lee2023electron, ponce2016epw} and \textsc{Wannier90}~\cite{pizzi2020wannier90} to compute electron-phonon coupling matrix elements and polarons~\cite{sio2019polarons}, and \textsc{BerkeleyGW} to perform GW/BSE calculations with finite exciton momentum~\cite{deslippe2012berkeleygw, hybertsen1986electron, rohlfing2000electron}. 
We set the kinetic energy cutoff for the dielectric matrix to 10~Ry, and include 5 valence bands and 195 conduction bands. To compute the self-energy, the COHSEX approximation is used; this choice yields a quasi-particle band gap of 14.7~eV, in good agreement with experiments~\cite{piacentini1976thermoreflectance}. The BSE kernel is constructed using 3 valence bands and 7 conduction bands; this choice is adequate to obtain a converged absorption spectrum near the band-edge, as can be seen by comparing Fig.~\ref{fig:lif_abs}(a) with Fig.~6 in Ref.~\cite{onida2002electronic}.
In addition, the software VESTA is used for the visualization of the crystal structures, charge densities, and displacement patterns~\cite{momma2011vesta}.

Equations (\ref{eqn:explrneqn})-(\ref{eqn:exphg}) are solved on uniform Brillouin zone grids with increasing density, from $4\times4\times4$ to $10\times10\times10$ points, corresponding to equivalent BvK supercells ranging from 128 to 2,000 atoms.
{We include 4 lowest-energy exciton bands when constructing the excitonic polaron Hamiltonian, which is sufficient to yield converged solutions [Fig. ~\ref{fig:lif_density}(a)]}
The initialization of the coefficients $A_{s\bmqq}$ is chosen such that $\Psi({\bf r}_e,{\bf r}_h)$ is already sufficiently localized, using the approach outlined by Sec.~\ref{sec:implication}. 
Specifically, we first set $A_{s\bmqq}$ to be a constant so that the normalization condition $N_p^{-1}\sum_{s\bmqq} \abs{A_{s\bmqq}}^2 = 1$ is satisfied, and we construct the exciton-phonon coupling matrix by imposing $g_{cc'\nu}(\bmk,\bmq)=0$. Then, we evaluate $B_{\bmqq \nu}$ using Eq.~(\ref{eqn:bmat}), and we diagonalize the matrix in Eq.~(\ref{eqn:explrneqn}); we repeat this process until convergence is achieved. The converged solution is subsequently employed to initialize a second run, where the complete exciton-phonon coupling matrix is used [i.e., without setting $g_{cc'\nu}(\bmk,\bmq)=0$], and a new iterative minimization is carried out.
{An estimate of the computational cost of our method is provided in App.~C.}

\subsubsection{Convergence behavior}

\begin{figure}
    \centering
    \includegraphics[width=0.5\textwidth]{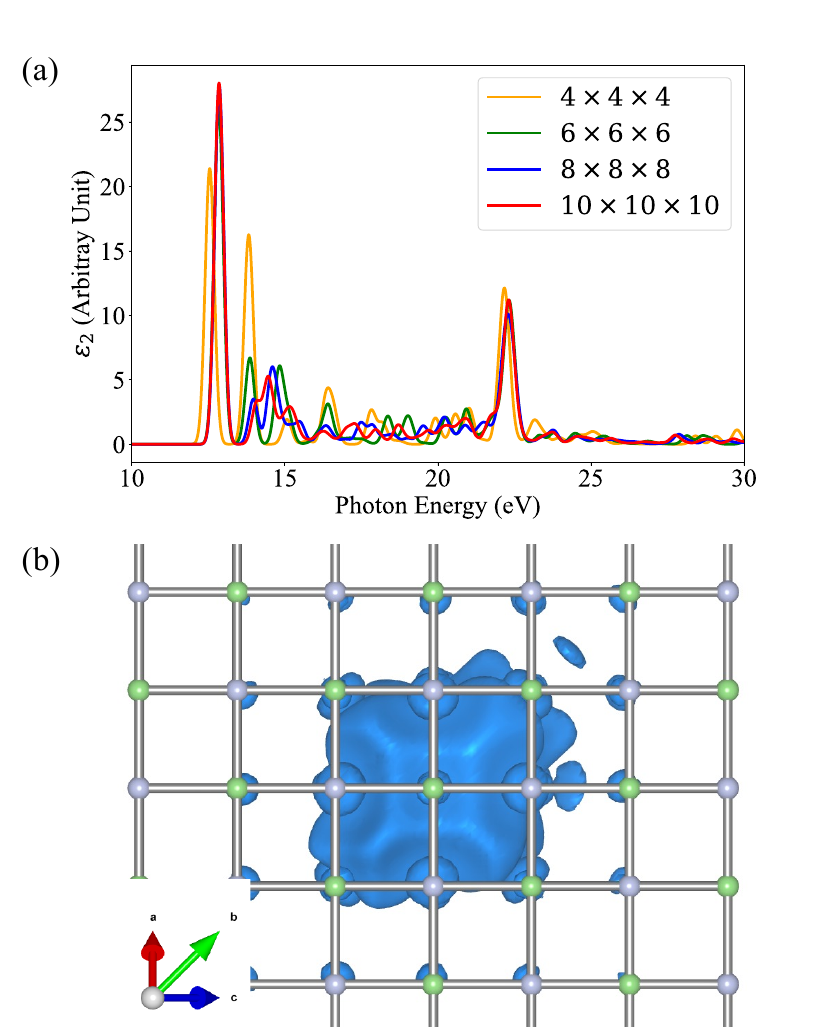}
    \caption{
    Optical absorption spectrum and exciton wavefunction of \ch{LiF}. (a) Imaginary part of the dielectric function $\epsilon_2$ of \ch{LiF}. With an $8\times8\times8$ Brillouin-zone grid, the lowest-energy peak is well converged. This peak provides the largest contribution to the formation of the excitonic polaron in LiF. (b) Electron charge density of the lowest-energy free exciton state, rendered in a $8\times8\times8$ supercell. The hole position is fixed near a fluorine atom. Li is in green, F is in silver. The electron charge density is relatively localized around the fixed hole.
    }
    \label{fig:lif_abs}
\end{figure}

Figure~\ref{fig:lif_abs}(a) shows the calculated BSE absorption spectrum of \ch{LiF} as a function of the Brillouin-zone mesh. We see that, in the vicinity of the absorption onset, the lineshape converges when a $8\times8\times8$ $\bmk$-grid is used (note that the green line for the $6\times6\times6$ $\bmk$-grid and the blue line for the $8\times8\times8$ $\bmk$-grid are hidden underneath the red line for the $10\times10\times10$ $\bmk$-grid); this finidng agrees well with previous studies on \ch{LiF}~\cite{onida2002electronic}. Based on this test, in Fig.~\ref{fig:lif_abs}(b) we plot the exciton electron density corresponding to an $8\times8\times8$ BvK supercell, with the hole position fixed on one of the fluorine atoms. In this panel, we see that the electron is fairly localized relative to the hole. This observation is consistent with the fact that a relatively coarse $\bmk$-grid yields a well-converged absorption spectrum. We emphasize that Fig.~\ref{fig:lif_abs}(b) shows \textit{relative} localization of the electron with respect to the hole, while the two-particle exciton wavefunction $\Omega^{s\bmqq}({\bf r}_e,{\bf r}_h)$ is fully \textit{delocalized} as well as translationally invariant.

Next, we investigate the exciton-phonon matrix elements that are central to Eqs.~(\ref{eqn:explrneqn}) and (\ref{eqn:bmat}). Since the first two exciton bands of \ch{LiF} are degenerate and isolated from the other bands [Fig.~\ref{fig:lif_density}(a), except for the $\Gamma$ point], in Fig.~\ref{fig:lif_exphg} we plot the gauge-invariant metric defined below:
\begin{align}
    \sqrt{\sum_{s,s'=1,2}\abs{\mathcal{G}_{ss'\nu}(\bmqq=0,\bmq)}^2 },
\end{align}
where $\nu$ is chosen to be LO mode that is non-degenerate for the most part of the Brillouin zone, and the path of $\bmq$ is a straight line from $\Gamma$ to $L$. 
We see that, as for the absorption spectrum in Fig.~\ref{fig:lif_abs}(b), the exciton-phonon matrix elements also converge when a $8\times8\times8$ grid is used for $\bmk$ points, $\bmq$ points, and $\bmqq$ points.

Interestingly, the momentum dependence of the exciton-phonon couplings computed from first principles and shown in Fig.~\ref{fig:lif_exphg} resembles the trend that we find for the Wannier-Fr\"ohlich model in Fig.~\ref{fig:analytical_exphg}. In particular, we find that the \textit{ab initio} matrix elements are also linear in $\bmq$ at small $\bmq$. However, unlike in the Wannier-Fr\"ohlich model which only considers one valence and one conduction bands, here we have 10 electronic bands in total. This difference might account  for the larger magnitude of the \textit{ab initio} exciton-phonon couplings shown in Fig.~\ref{fig:lif_exphg} as compared to the model calculation of Fig.~\ref{fig:analytical_exphg}. This stronger coupling may favor the formation of the excitonic polaron in LiF.

\begin{figure}
    \centering
    \includegraphics[width=0.5\textwidth]{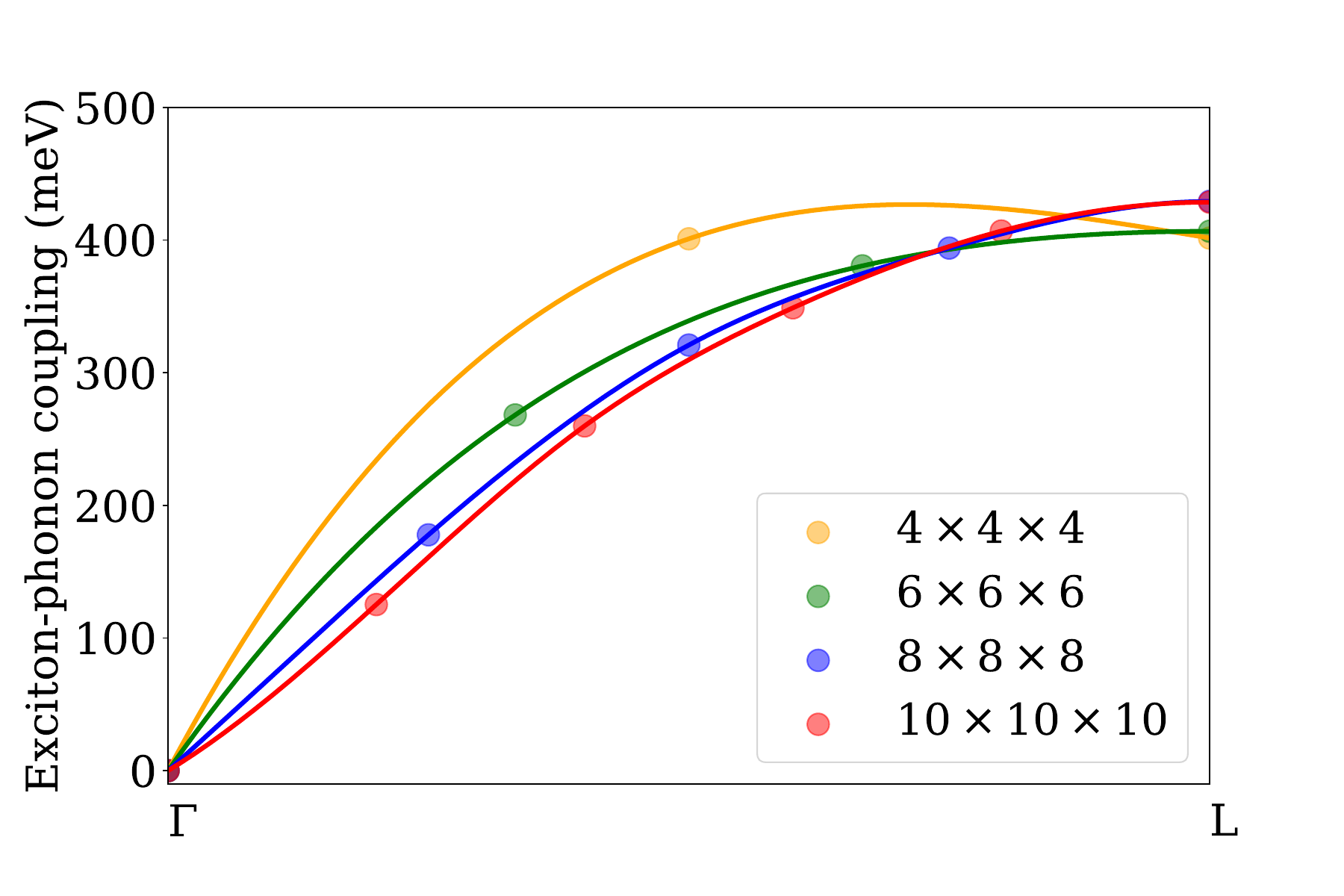}
    \caption{
    Exciton-phonon coupling matrix elements of \ch{LiF}. As for the absorption spectrum, the exciton-phonon coupling matrix elements also converges with an $8 \times 8 \times 8$ Brillouin-zone grid. The behavior of these matrix elements resmbles the Fr\"{o}hlich type exciton-phonon interaction shown in Fig.~\ref{fig:analytical_exphg}, although here the matrix elements are 2-3 times higher. This comparison suggests that the Wannier-Fr\"{o}hlich model captures qualitative trends but does not carry quantitative accuracy. The discs are our calculations, the lines are guides to the eye.}
    \label{fig:lif_exphg}
\end{figure}

\subsubsection{Formation energy}

Once the excitonic polaron equations [Eqs.~(\ref{eqn:explrneqn})-(\ref{eqn:exphg})] are solved,  the formation energy of the excitonic polaron can be computed from $\Delta E^{f}_{\mathrm{xp}} = E_{\mathrm{xp}} - E^0_{\mathrm{BSE}}$, where $E_{\mathrm{xp}}$ is given by Eq.~(\ref{eqn:exciton_etot}), and $E^0_{\mathrm{BSE}}$ the lowest BSE eigenvalue for the undistorted structure. The results are show in in Fig.~\ref{fig:lif_density}(b). 
First, we find that the formation energy of the excitonic polaron is negative, meaning that the polaron state is stable as compared to free excitons. In addition, we notice that the formation energy decreases when a denser Brillouin zone grid is used. The extrapolation to infinite BvK supercell ($N_p \rightarrow \infty$) corresponds to a fully isolated excitonic polaron, in analogy with the case of electron and hole polarons~\cite{sio2019ab,sio2019polarons}.

In Tab.~\ref{tab:energies}, we show that the formation energy of the excitonic polaron in \ch{LiF} lies between the energy of electron polaron and that of the hole polaron.
This finding can be understood from Eq.~(\ref{eqn_wf:etot_density}), where the formation of the excitonic polaron could be viewed as the combination of the electron polaron and hole polaron attracting each other, thus making the highly localized hole polaron stabilize the much more diffuse electron polaron.

In the companion manuscript \cite{dai2023explrn_prl}, we show that for another compound, \ch{Cs2ZrBr6}, both the electron polaron and the hole polaron are highly localized. In that case the excitonic polaron is less stable than both the electron and the hole polaron as a result of the cancellation of the two charge densities, which reduces the electrostatic interaction with the ionic lattice.


\begin{table}[]
\begin{tabular}{*{3}{>{\centering\arraybackslash}m{0.1\textwidth}} *{2}{>{\centering\arraybackslash}m{0.07\textwidth}} }
\toprule
\toprule
\multicolumn{3}{c}{Polaron formation energy} & \multirow{2}{*}{$\mathrm{E_{gap}^{GW}}$} & \multirow{2}{*}{$\mathrm{E_{b}^{ex}}$} \\ \cmidrule(r){1-3}
Electron       & Hole       & Excitonic      &                     &                     \\ \midrule
-231 meV       & -1980 meV  & -461 meV       & 14.7 eV             & 1.88 eV             \\
\bottomrule
\bottomrule
\end{tabular}
\caption{Formation energies of the electron polaron, hole polaron, and excitonic polaron in \ch{LiF}. These values correspond to the limit of infinite supercell. For completeness, we also report the GW quasi-particle band gap $\mathrm{E_{gap}^{GW}}$ and the binding energy of the lowest-energy exciton $\mathrm{E_{b}^{ex}}$. 
}
\label{tab:energies}
\end{table}

\begin{figure*}
    \centering
    \includegraphics[width=1.0\textwidth]{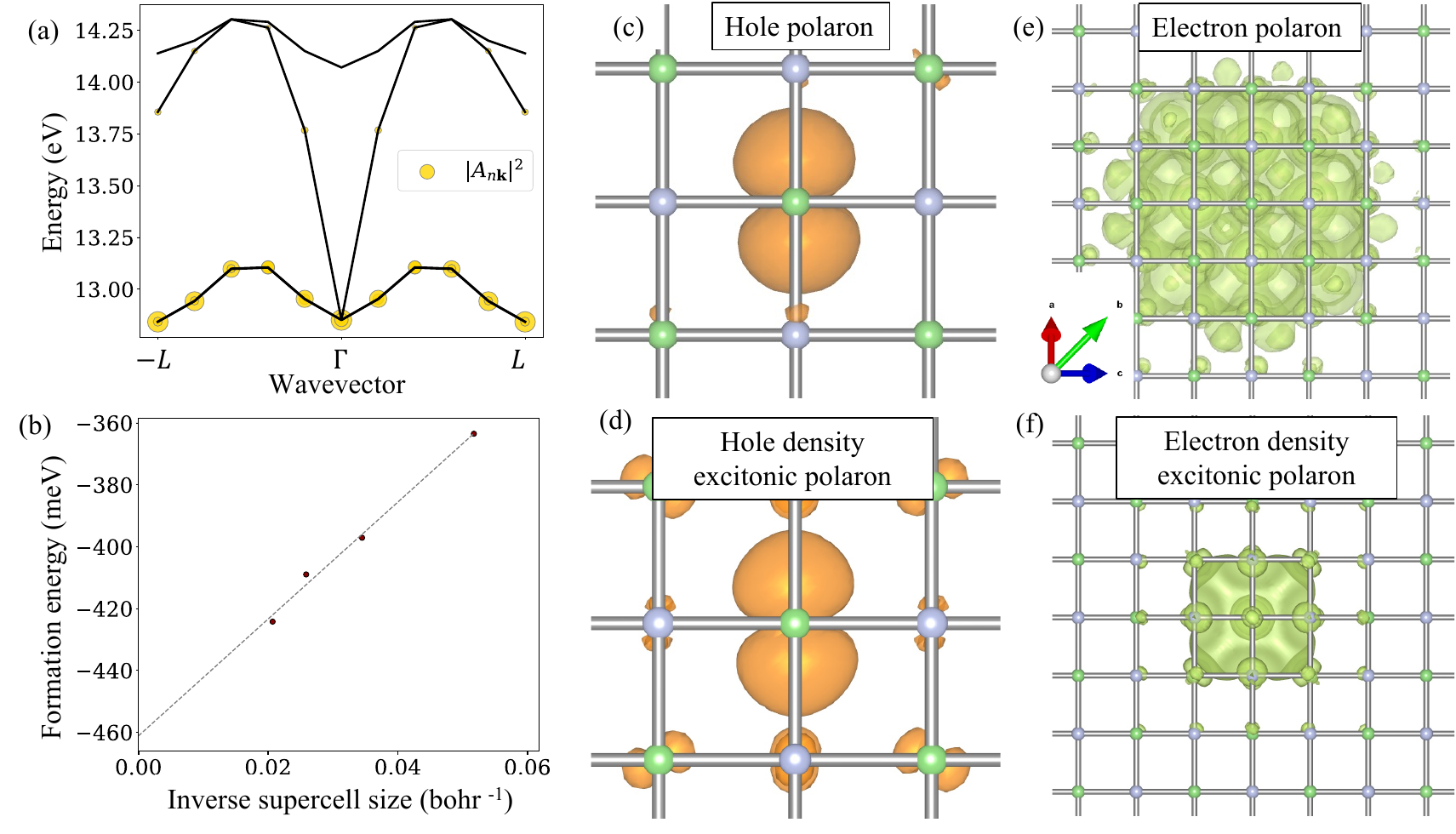}
    \caption{Exciton band structure, convergence of the formation energy of excitonic polarons with Brillouin-zone sampling, and densities of polarons and excitonic polaron in \ch{LiF}. (a) 
    The contribution of each free exciton state to the formation of excitonic polarons. 
    The black lines denote the exciton band structure of \ch{LiF}, and the yellow discs are the  contribution $\abs{A_{s\bmqq}}^2$ of each exciton to the formation of the excitonic polaron. The most significant contributions come from the three lowest exciton bands, which correspond to the lowest-energy peak in the absorption spectrum shown in Fig.~\ref{fig:lif_abs}(a). (b) Formation energy of the excitonic polaron as a function of the inverse supercell size. The infinite supercell limit corresponds to a fully isolated excitonic polaron. The formation energy extrapolated to this limit is $-461$~meV. (c) Charge density of the hole polaron (orange isosurface). (d) Hole density of the excitonic polaron (orange isosurface). (e) Charge density of the large electron polaron (green isosurface). (f) Electron density of the excitonic polaron (green isosurface). In (c)-(f), Li atoms are in green, F atoms are in silver. 
}
    \label{fig:lif_density}
\end{figure*}

\subsubsection{Charge densities}

In the case of free excitons, the common practice when plotting exciton charge densities is to fix the position of the hole (or the electron), and then plot the electron (or the hole) density. This is shown in Fig.~\ref{fig:lif_abs}(a) for LiF. The mathematical expression describing this procedure is:
\begin{align}\label{eq.tmp.3}
    n^{s\bmqq}_e(\bmr_e; \bmr_h^0) = \abs{\Omega_{s\bmqq}(\bmr_e,\bmr_h^0)}^2,
\end{align}
which gives the electron density of the exciton $s\bmqq$ when the hole position is fixed at $\bmr_h^0$.
Now, if we integrate $n^{s\bmqq}_e(\bmr_e; \bmr_h^0)$ over the hole position $\bmr_h^0$, the resulting electron density becomes fully delocalized and exhibits the periodicity of the lattice. This property can easily be seen by using Eq.~(\ref{eqn:basis_transform}) inside Eq.~\eqref{eq.tmp.3}.

In the case of excitonic polarons, the lattice distortion favors exciton localization, thus breaking the lattice periodicity. To demonstrate this point, we evaluate and visualize the electron and hole charge densities of the excitonic polaron using the following expressions:
\begin{align}
    \label{eqn:plot_charge_density}
    &n_e(\bmr_e) = \int{d\bmr_h} \abs{\Psi(\bmr_e, \bmr_h)}^2
    =
    \frac{1}{N_p} \sum_{v\bmk}
    \abs{L_{v\bmk} (\bmr_e)}^2,
     \\
    \label{eqn:plot_charge_density2}
    &n_h(\bmr_h) = \int{d\bmr_e} \abs{\Psi(\bmr_e, \bmr_h)}^2
    =  
    \frac{1}{N_p} \sum_{c\bmk}
    \abs{L_{c\bmk} (\bmr_h)}^2,
\end{align}
where the auxiliary functions $L_{v\bmk}$ and $L_{c\bmk}$ are given by:
\begin{align}
\label{eqn:l_functions}
    &L_{v\bmk} (\bmr_e) 
    =
    \sum_{s\bmqq} \sum_c A_{s\bmqq} a_{\vck}^{s\bmqq}
    \psi_{c\bmk+\bmqq}(\bmr_e),
    \nonumber \\
    &L_{c\bmk} (\bmr_h) 
    =
    \sum_{s\bmqq} \sum_v A_{s\bmqq} a_{\vck-\bmqq}^{s\bmqq}
    \psi^*_{v\bmk-\bmqq}(\bmr_h).
\end{align}
Since these functions are linear combinations of Bloch states with different crystal momenta, they do not possess lattice periodicity in general. Therefore, the charge densities obtained from Eqs.~\eqref{eqn:plot_charge_density}-\eqref{eqn:plot_charge_density2} are allowed to be localized within the BvK supercell.

We also emphasize that the localization of the excitonic polaron as described by Eqs.~\eqref{eqn:plot_charge_density}-\eqref{eqn:plot_charge_density2} is closely related to the approximation of classical nuclei that underpins all DFT, DFPT, and GW/BSE calculations.
{
In this context, the breaking of lattice-periodicity means that for a fixed displacement pattern, the corresponding excitonic polaron wave function will break the lattice periodicity.
Shifting the displacement pattern by a lattice vector $\bmrr$ will yield another localized excitonic polaron wave function with the same shape, but it is shifted by $\bmrr$ as well. 
}
In a more sophisticated, quantum treatment of atomic displacements, the electron and hole charge densities would be localized with respect to each other and with respect to the distortion of the atomic lattice, but the composite excitation consisting of electron, hole, and phonon cloud would still be a delocalized entity in agreement with Bloch's theorem. Therefore, to be more precise, the excitonic polaron discussed below is what might be called a ``pinned'' excitonic polaron. 
{
We note that this loss of translational invariance is entirely analogous to what happens in calculations of charged polarons~\cite{sio2019ab, sio2022unified, lafuente2022ab, lafuente2022unified}.
Restoring full translational invariance would require one to consider a Green’s function that includes both electronic and vibrational degrees of freedom, as it is done, for example, in Diagrammatic Monte Carlo calculations~\cite{prokof1998polaron, mishchenko2000diagrammatic}.}

The charge densities for the electron polaron, hole polaron, and excitonic polaron in \ch{LiF} are shown in Fig.~\ref{fig:lif_density}(c)-(f). 
We find that the hole density of the excitonic polaron [Fig.~\ref{fig:lif_density}(d)] has the shape of $p$-orbitals, and it is largely localized around a fluorine atom. 
{
There are other degenerate solutions in the same unit cell, which correspond to $p$-orbitals oriented toward the other Cartesian directions and are accompanied by different atomic displacement patterns. This multiplicity is a consequence of the fact that our formalism described “pinned” excitonic polarons.
}
Furthermore, the hole density of the excitonic polaron is very similar to that of the hole polaron [Fig.~\ref{fig:lif_density}(c)].
On the other hand, the electron density of the excitonic polaron [Fig.~\ref{fig:lif_density}(f)] is similar in shape to the electron polaron [Fig.~\ref{fig:lif_density}(e)], but it is considerably more localized, spanning only a couple of unit cells.

The similarity between the charge densities of the electron and hole polaron and the charge densities of the excitonic polaron shown in Fig.~\ref{fig:lif_density}(c)-(f) suggests that the formation of the excitonic polaron in \ch{LiF} might be viewed as a two-step process: (i) the formation of an electron polaron and a hole polaron which do not interact with each other; (ii) the formation of the excitonic polaron as a result of the mutual Coulomb attraction of these polarons. In this latter step, the small hole polaron acts a pinning center for the large electron polaron. In this case, it is not necessary to fix the hole center as in the visualization of the free exciton in Fig.~\ref{fig:lif_abs}(b). 
{The similarity with Fig. \ref{fig:lif_density}(f) indicates that the binding of the exciton is so strong that the lattice distortion only slightly influences the mutual interaction between electrons and holes.}
The present picture is fully consistent with the analysis of the mechanism of formation of the excitonic polaron in the Wannier exciton model presented in Sec.~\ref{sec:etot_transition}.

\begin{figure*}
    \centering
    \includegraphics[width=1.0\textwidth]{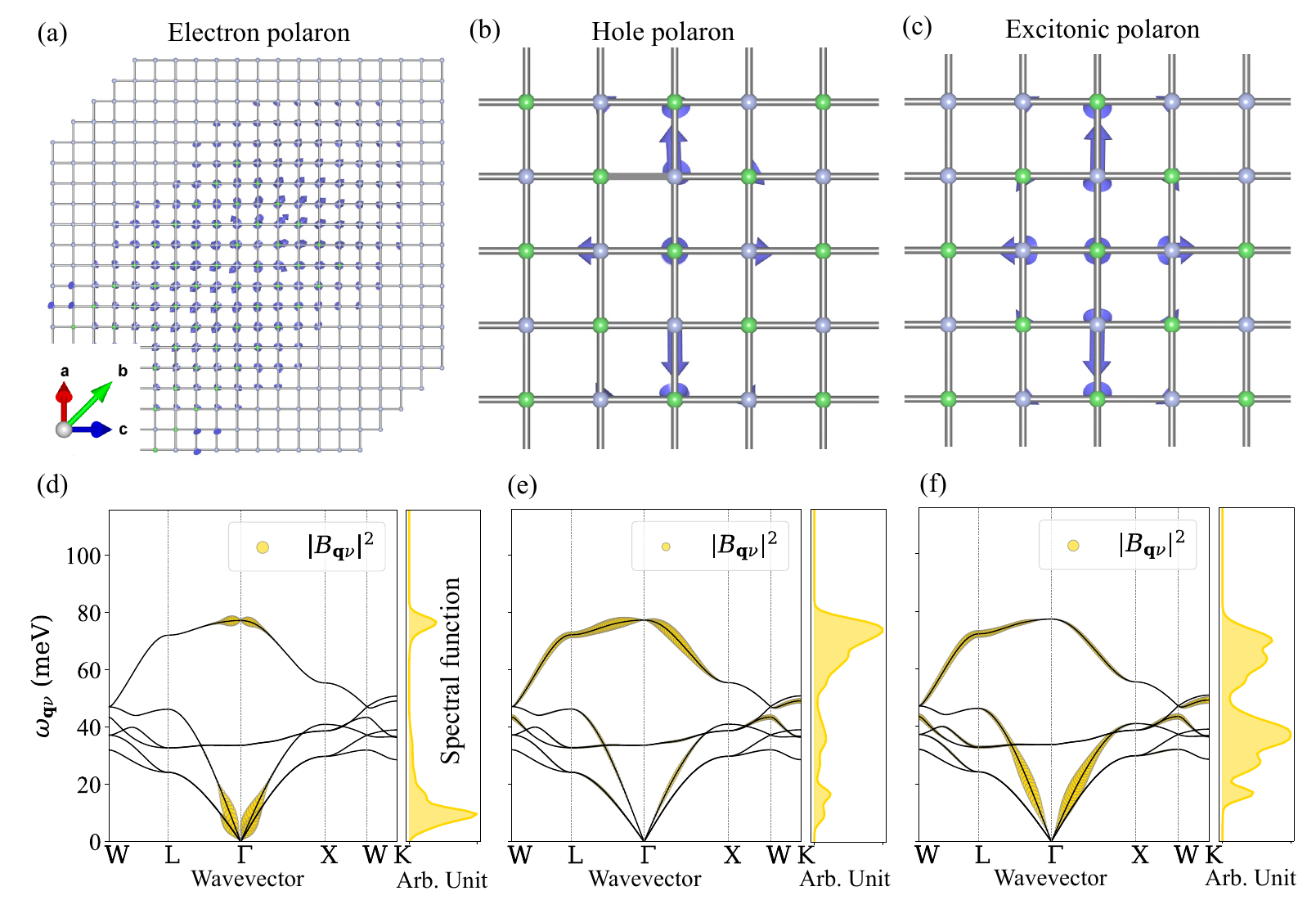}
    \caption{Displacement patterns of polarons and excitonic polaron in \ch{LiF}, and corresponding phonon contributions. In (a)-(c), Li and F atoms are in green and silver, respectively, and purple arrows represent atomic displacements. Only significant displacements are displayed for clarity. In (d)-(f), black lines represent the phonon dispersions, and the yellow discs represent the phonon contribution $\abs{B_{\bmq \nu}}^2$ to the formation of polarons. {The shadowed area on the right in each panel represents the spectral function $B^2(E).$ They are normalized so that the top of the range coincides with the highest peak in each case. } (a) Atomic displacements associated with the electron polaron. (b) Atomic displacements associated with the hole polaron. (c) Atomic displacements associated with the excitonic polaron. (d) Contributions of individual phonons to the formation of the electron polaron. The acoustic branches and the LO modes around the zone center contribute the most. (e) Contributions of individual phonons to the hole polaron. (f) Contributions of individual phonons to the formation of the excitonic polaron.}
    \label{fig:lif_disp}
\end{figure*}

Figure~\ref{fig:lif_density}(a) shows the contribution of each exciton state of the undistorted structure to the formation of the excitonic polaron, as given by the weights $|A_{s\bmqq}|^2$. We find that the the lowest exciton band carries the largest contribution, with smaller contributions from the next two bands. The fact that only the low-lying exciton bands contribute to the formation of the excitonic polaron provides \textit{a posteriori} support to our choice of solving the excitonic polaron equations in the exciton basis [Eqs.~(\ref{eqn:explrneqn})-(\ref{eqn:bmat})] rather than in the transition basis [Eqs.~(\ref{eqn_rs:polaron_eqn_final})-(\ref{eqn_rs:g_transition})]. In Fig.~\ref{fig:lif_density}(a), we also see that the coefficients $|A_{s\bmqq}|^2$ are significant throughout the entire Brillouin zone. This ``delocalization'' in reciprocal space is consistent with our observation of strong real-space localization of the excitonic polaron. 

\subsubsection{Displacement patterns}

Figure~\ref{fig:lif_disp} shows the atomic displacements associated with the electron polaron, the hole polaron, and the excitonic polaron in LiF. These displacements are evaluated using Eq.~(\ref{eqn:displacement}). As expected, the displacement pattern of the excitonic polaron [Fig.~\ref{fig:lif_disp}(c)] is centered around the hole charge density of the excitonic polaron, and it is highly localized. This pattern is very similar to what we find for the hole polaron in Fig.~\ref{fig:lif_disp}(b), but differs considerably from the displacements obtained for the electron polaron which is more delocalized [Fig.~\ref{fig:lif_disp}(a)].

A more detailed comparison between the atomic displacement patters of electron polaron, hole polaron, and excitonic polarons in LiF can be performed by inspecting the $B_{\bmq\nu}$ coefficients. Indeed, according to Eq.~(\ref{eqn:displacement}), these coefficients can be thought of as the contributions from individual phonons to the formation of these localized quasiparticles. Figures~\ref{fig:lif_disp}(d) and (e) show how the electron polaron is dominated by the coupling with long-wavelength modes, and in particular the LO mode; while the hole polaron draws weight from the entire Brillouin zone. These trends are consistent with previous work on polarons in LiF~\cite{sio2019polarons}, and indicate that this compound hosts Fr\"ohlich electron polarons and Holstein hole polarons, respectively.
In the case of the excitonic polaron, Fig.~\ref{fig:lif_disp}(f) shows that the contribution from the LO mode around the zone center is much smaller than for the electron and hole polarons. This effect can be rationalized in terms of the qualitative difference between exciton-phonon and electron-phonon couplings to the LO mode: while Fr\"{o}hlich electron-phonon coupling goes as $1/|{\bf q}|$ for $\bmq \rightarrow 0$, the exciton-phonon coupling is not singular and goes as $|{\bf q}|$ [Fig.~\ref{fig:analytical_exphg} and Fig.~\ref{fig:lif_exphg}], therefore the contribution of long-wavelength polar phonons is less significant in the case of excitonic polarons.
{Furthermore, we notice the significant contribution coming from the acoustic modes in Fig.~\ref{fig:lif_disp}(f), whose contribution is typically neglected in the model Hamiltonian approaches. 
In fact, by manually removing the contribution from the acoustic modes, the polaron formation energy will be reduced from $-409$ meV to $-206$ meV (on an $8 \times 8 \times 8$ grid). 
This clearly demonstrates the necessity of performing full {\it ab initio} calculations to describe excitonic polarons in real materials.}

Another interesting behavior that emerges from Figs.~\ref{fig:lif_disp} is that the coefficients $B_{\bmq \nu}$ of the excitonic polaron [Figs.~\ref{fig:lif_disp}(f)] resemble a superposition of the coefficients $B_{\bmq \nu}$ that we find for the electron polaron [Figs.~\ref{fig:lif_disp}(d)] and the hole polaron [Figs.~\ref{fig:lif_disp}(e)].
{This can be further illustrated by the spectral function in the right panels of Figs.~\ref{fig:lif_disp}(d)-(f), defined as:
\begin{align}
    B^2(E)=\frac{1}{N_p} \sum_{\bmq\nu}\abs{B_{\bmq\nu}}^2\delta(E-\hbar\omega_{\bmq\nu}).
\end{align} 
We find that the hole polaron and the excitonic polaron spectral functions show similar peak strengths for the LO modes, while the electron polaron and the excitonic polaron spectral functions shows similar peak strengths for the acoustic modes.
Since acoustic modes are mainly responsible for elastic deformation while optical modes usually induce out-of-phase motion, the similarity of the spectral functions in Figs.~\ref{fig:lif_disp}(e) and (f) is consistent with the similar displacement patterns for the hole polaron and the excitonic polaron in Figs.~\ref{fig:lif_disp}(b) and (c).}
This observation further supports the conceptual scenario whereby the excitonic polaron can be thought of as if being formed from one electron polaron and one hole polaron bound together by their mutual Coulomb attraction.

\section{Summary and outlook}
\label{sec:summary}

In this work we presented a first-principles theory of excitonic polarons that combines the Bethe-Salpeter equation approach for excitons with density-functional perturbation theory for phonons. Our theory directly yields the energetics, wavefunctions, and atomic displacements of excitonic polarons, without requiring supercells. The only ingredients needed for performing these calculations are the band structures or electrons, phonons, and excitons, and the electron-phonon matrix elements. All these quantities can be computed from calculations in the crystal unit cell.

In the present approach, the search for excitonic polarons is formulated as a variational minimization of the excited state total energy, which consists of the DFT total energy and the BSE excitation energy. Explicit supercell calculations are avoided by expressing the excitonic polaron wavefunction as a linear combination of finite-momentum excitons. This strategy leads to a nonlinear system of two coupled equations for the exciton polaron wave function and the associated atomic displacements, which is reminiscent of the polaron equations introduced for charged particles~\cite{sio2019ab}.

We have identified two possible sets of equations to obtain excitonic polarons, one set in the transition basis, and one in the exciton basis. The exciton basis formulation is most suited for \textit{ab initio} calculations. The transition basis formulation is useful to make contact with model Hamiltonians. For example, using the transition basis we have shown that a Wannier exciton model with Fr\"{o}hlich electron-phonon interactions leads to an excitonic polaron equation that is similar to the Landau-Pekar equation for polarons. The analysis of this simplified model suggests that the excitonic polaron can be thought of as an excitation resulting from the binding of an electron polaron and a hole polaron via their mutual Coulomb attraction. The analysis of the Wannier exciton with Fr\"{o}hlich or Holstein electron-phonon interactions also allowed us to identify two general criteria that must be met for excitonic polarons to form: (i) for excitonic polarons dominated by Fr\"{o}hlich couplings, the electron and hole effective masses should differ significantly; (ii) for excitonic polarons dominated by Holstein couplings, there must be a large difference between the short-range electron-phonon interactions and hole-phonon interactions.

We applied this method to lithium fluoride, a prototypical material that hosts small hole polarons and large electron polarons. Our first-principles calculations reveal that the exciton-phonon coupling matrix elements in a polar insulator decrease linearly with the exciton momentum at long wavelength. This behavior is qualitatively different from the well-known divergence of the Fr\"{o}hlich electron-phonon interaction in three-dimensional materials at long wavelength, and is reflected in the small contribution of phonon modes around the zone center to the excitonic polaron. We also find that the hole charge density and atomic displacements of the excitonic polaron in LiF strongly resemble those of the hole polaron, suggesting that the localization of the excitonic polaron is largely dictated by the most localized quasiparticle among the electron and the hole polaron.
To quantify which exciton states, phonon modes, and exciton-phonon couplings contribute the most to the formation of the excitonic polaron, we performed a spectral analysis of the wavefunction and displacements in LiF, and we found that only the three lowest-energy exciton bands play a significant role in the formation of the excitonic polaron.

In the companion manuscript~\cite{dai2023explrn_prl}, we apply the same formalism presented here to a more complex material, the vacancy-ordered double perovskite \ch{Cs2ZrBr6}. In that case, we observe similar relations between the electron polaron, the hole polaron, and the excitonic polaron as discussed here for LiF, suggesting that these interrelations might be universal features in the physics of polarons and excitonic polarons.

The present development opens several possible avenues for future work. Firstly, it will be important to extend these calculatios to a broad materials dataset to map out the properties of excitonic polarons across diverse materials families. Secondly, it will be important to connect this methodology to experimental measurements of Stokes shifts between absorption and luminescence spectra. Thirdly, applications of the present methodology to the case of two-dimensional materials will be of considerable interest. Lastly, re-deriving the present formalism starting from a general field-theoretic approach would be highly desirable. For example, work generalizing the many-body theory of polarons of Ref.~\citenum{lafuente2022ab} to the case of excitonic polarons would be useful.

We hope that this work will serve as the basis for systematic {\it ab initio} calculations of excitonic polarons in real materials, and that it will form the starting point for further investigations of the physics of self-localization in condensed matter systems.

\begin{acknowledgments}
This research is primarily supported by the National Science Foundation, Office of Advanced Cyberinfrastructure and Division of Materials Research under Grant No. 2103991 of the Cyberinfrastructure for Sustained Scientific Innovation program, and the NSF Characteristic Science Applications for the Leadership Class Computing Facility program under Grant No. 2139536 (development of the excitonic polaron module, calculations, interoperability). This research was also supported by the Computational Materials Sciences Program funded by the US Department of Energy, Office of Science, Basic Energy Sciences, under award no. DE-SC0020129 (development of the polaron module). This research used resources of the National Energy Research Scientific Computing Center and the Argonne Leadership Computing Facility, which are DOE Office of Science User Facilities supported by the Office of Science of the U.S. Department of Energy, under Contracts No. DE-AC02-05CH11231 and DE-AC02-06CH11357, respectively. The authors acknowledge the Texas Advanced Computing Center (TACC) at The University of Texas at Austin for providing access to Frontera, Lonestar6, and Texascale Days, that have contributed to the research results reported within this paper (http://www.tacc.utexas.edu).
\end{acknowledgments}

\newpage
\appendix
\begin{widetext}
\section{Complete formulas for Eq.~(\ref{eqn:ene_ph_froe})}
\label{app:full_froe}

The terms $t_1$-$t_{10}$ required in Eq.~(\ref{eqn:ene_ph_froe}) are given below:
\begin{align}
    t_1&=a^6 b^6 a_0^9 (5 + 4ab),
\\
    t_2&=7 a^5 b^5 a_0^8 r_p (5+4ab),
\\
    t_3&= a^4 b^4 a_0^7 r_p^2 (101 a^4 + 519 a^3 b + 848 a^2 b^2 + 519 a b^3 + 101 b^4),
\\
    t_4&=7 a^3 b^3 a_0^6 r_p^3 (1+2ab) (21 + 4ab),
\\
    t_5&=a^2 b^2 a_0^5 r_p^4 (101 a^6 + 1149 a^5 b + 3858 a^4 b^2 + 5632 a^3 b^3 + 3858 a^2 b^4 + 1149 a b^5 + 101 b^6),
\\
    t_6&=7 a b a_0^4 r_p^5 (5a^6 + 88a^5b + 368a^4 b^2 + 574 a^3 b^3 + 368 a^2 b^4 + 88 a b^5 + 5 b^6),
\\
    t_7&=a_0^3 r_p^6 (5a^8 + 193a^7b + 1404a^6b^2 + 4020a^5b^3 + 5612a^4b^4 + 4020a^3b^5 + 1404a^2b^6 + 193ab^7 + 5b^8),
\\
    t_8&=4a_0^2r_p^7(6+37ab),
\\
    t_9&=28a_0r_p^8,
\\
    t_{10}&=4r_p^9.
\end{align}

\section{Complete formulas for Eq.~(\ref{eqn:ene_ph_hols})}
\label{app:full_hols}

The integral in Eq.~(\ref{eqn:ene_ph_hols}) can be written as:
\begin{equation}
    E^{\mathrm{H}}_{\rm ph} 
    =
    -\frac{1}{2 \hbar \omega_{\mathrm{LO}}}
    \frac{1}{\pi^2} 
    I^{\mathrm{H}},
\end{equation}
where the term $I^{\mathrm{H}}$ has been evaluated using Mathematica~\cite{Mathematica}. This term is given by:
\begin{align}
    &I^{\mathrm{H}}
    =
    \left[{4 (a a_0+r_p)^7 (a_0 b+r_p)^7}\right]^{-1} \pi
    \nonumber \\
    &\times
    \Bigg(a_0^7 g_c^2 \left(a_0^4 b^4+7 a_0^3 r_p b^3+17 a_0^2 r_p^2 b^2+7 a_0 r_p^3 b+r_p^4\right) a^{10}
    \nonumber \\
    &+a_0^6 g_c^2 \left(3 a_0^5 b^5+28 a_0^4 r_p b^4+100 a_0^3 r_p^2 b^3+140 a_0^2 r_p^3 b^2+52 a_0 r_p^4 b+7 r_p^5\right) a^9
    \nonumber \\
    &+3 a_0^5 g_c^2 \left(a_0^6 b^6+14 a_0^5 r_p b^5+73 a_0^4 r_p^2 b^4+175 a_0^3 r_p^3 b^3+169 a_0^2 r_p^4 b^2+56 a_0 r_p^5 b+7 r_p^6\right) a^8
    \nonumber \\
    &+a_0^4 \big[a_0^7 \left(g_c^2-16 g_v g_c+g_v^2\right) b^7+7 a_0^6 \left(4 g_c^2-16 g_v g_c+g_v^2\right) r_p b^6+a_0^5 \left(227 g_c^2-336 g_v g_c+21 g_v^2\right) r_p^2 b^5
    \nonumber \\
    &+35 a_0^4 \left(24 g_c^2-16 g_v g_c+g_v^2\right) r_p^3 b^4+a_0^3 \left(1464 g_c^2-560 g_v g_c+35 g_v^2\right) r_p^4 b^3+7 a_0^2 \left(151 g_c^2-48 g_v g_c+3 g_v^2\right) r_p^5 b^2
    \nonumber \\
    &+7 a_0 \left(44 g_c^2-16 g_v g_c+g_v^2\right) r_p^6 b+\left(35 g_c^2-16 g_v g_c+g_v^2\right) r_p^7\big] a^7
    \nonumber \\
    &+a_0^3 \big[3 a_0^8 g_v^2 b^8+7 a_0^7 \left(g_c^2-16 g_v g_c+4 g_v^2\right) r_p b^7+16 a_0^6 \left(7 g_c^2-45 g_v g_c+7 g_v^2\right) r_p^2 b^6+7 a_0^5 \left(95 g_c^2-272 g_v g_c+36 g_v^2\right) r_p^3 b^5
    \nonumber \\
    &+2 a_0^4 \left(945 g_c^2-1368 g_v g_c+175 g_v^2\right) r_p^4 b^4+14 a_0^3 \left(177 g_c^2-167 g_v g_c+22 g_v^2\right) r_p^5 b^3+a_0^2 \left(1393 g_c^2-1198 g_v g_c+168 g_v^2\right) r_p^6 b^2
    \nonumber \\
    &+2 a_0 \left(175 g_c^2-171 g_v g_c+26 g_v^2\right) r_p^7 b+7 \left(5 g_c^2-6 g_v g_c+g_v^2\right) r_p^8\big] a^6
    \nonumber \\
    &+a_0^2 (a_0 b+r_p) \big[3 a_0^8 g_v^2 b^8+39 a_0^7 g_v^2 r_p b^7+a_0^6 \left(21 g_c^2-336 g_v g_c+188 g_v^2\right) r_p^2 b^6+a_0^5 \left(231 g_c^2-1568 g_v g_c+477 g_v^2\right) r_p^3 b^5
    \nonumber \\
    &+6 a_0^4 \left(161 g_c^2-472 g_v g_c+120 g_v^2\right) r_p^4 b^4+a_0^3 \left(1722 g_c^2-2614 g_v g_c+673 g_v^2\right) r_p^5 b^3+2 a_0^2 \left(483 g_c^2-665 g_v g_c+192 g_v^2\right) r_p^6 b^2
    \nonumber \\
    &+3 a_0 \left(77 g_c^2-118 g_v g_c+41 g_v^2\right) r_p^7 b+\left(21 g_c^2-38 g_v g_c+17 g_v^2\right) r_p^8\big] a^5
    \nonumber \\
    &+a_0 \big[a_0^{10} g_v^2 b^{10}+28 a_0^9 g_v^2 r_p b^9+219 a_0^8 g_v^2 r_p^2 b^8+35 a_0^7 \left(g_c^2-16 g_v g_c+24 g_v^2\right) r_p^3 b^7+2 a_0^6 \left(175 g_c^2-1368 g_v g_c+945 g_v^2\right) r_p^4 b^6
    \nonumber \\
    &+7 a_0^5 \left(199 g_c^2-778 g_v g_c+384 g_v^2\right) r_p^5 b^5+2 a_0^4 \left(1239 g_c^2-2866 g_v g_c+1239 g_v^2\right) r_p^6 b^4+2 a_0^3 \left(945 g_c^2-1727 g_v g_c+732 g_v^2\right) r_p^7 b^3
    \nonumber \\
    &+35 a_0^2 \left(19 g_c^2-34 g_v g_c+15 g_v^2\right) r_p^8 b^2+4 a_0 \left(28 g_c^2-53 g_v g_c+25 g_v^2\right) r_p^9 b+7 (g_c-g_v)^2 r_p^{10}\big] a^4
    \nonumber \\
    &+r_p \big[7 a_0^{10} g_v^2 b^{10}+100 a_0^9 g_v^2 r_p b^9+525 a_0^8 g_v^2 r_p^2 b^8+a_0^7 \left(35 g_c^2-560 g_v g_c+1464 g_v^2\right) r_p^3 b^7
    \nonumber \\
    &+14 a_0^6 \left(22 g_c^2-167 g_v g_c+177 g_v^2\right) r_p^4 b^6+a_0^5 \left(1057 g_c^2-3944 g_v g_c+2688 g_v^2\right) r_p^5 b^5+2 a_0^4 \left(732 g_c^2-1727 g_v g_c+945 g_v^2\right) r_p^6 b^4
    \nonumber \\
    &+840 a_0^3 (g_c-g_v)^2 r_p^7 b^3+a_0^2 \left(227 g_c^2-446 g_v g_c+219 g_v^2\right) r_p^8 b^2+28 a_0 (g_c-g_v)^2 r_p^9 b+(g_c-g_v)^2 r_p^{10}\big] a^3
    \nonumber \\
    &+b r_p^2 \big[17 a_0^9 g_v^2 b^9+140 a_0^8 g_v^2 r_p b^8+507 a_0^7 g_v^2 r_p^2 b^7+7 a_0^6 \left(3 g_c^2-48 g_v g_c+151 g_v^2\right) r_p^3 b^6
    \nonumber \\
    &+a_0^5 \left(168 g_c^2-1198 g_v g_c+1393 g_v^2\right) r_p^4 b^5+a_0^4 \left(507 g_c^2-1684 g_v g_c+1197 g_v^2\right) r_p^5 b^4+35 a_0^3 \left(15 g_c^2-34 g_v g_c+19 g_v^2\right) r_p^6 b^3
    \nonumber \\
    &+a_0^2 \left(219 g_c^2-446 g_v g_c+227 g_v^2\right) r_p^7 b^2+42 a_0 (g_c-g_v)^2 r_p^8 b+3 (g_c-g_v)^2 r_p^9\big] a^2
    \nonumber \\
    &+b^2 r_p^3 (7 a_0 b+3 r_p) \big[a_0^7 g_v^2 b^7+7 a_0^6 g_v^2 r_p b^6+21 a_0^5 g_v^2 r_p^2 b^5+a_0^4 \left(g_c^2-16 g_v g_c+35 g_v^2\right) r_p^3 b^4+7 a_0^3 \left(g_c^2-6 g_v g_c+5 g_v^2\right) r_p^4 b^3
    \nonumber \\
    &+a_0^2 \left(17 g_c^2-38 g_v g_c+21 g_v^2\right) r_p^5 b^2+7 a_0 (g_c-g_v)^2 r_p^6 b+(g_c-g_v)^2 r_p^7\big] a
    \nonumber \\
    &+b^3 r_p^4 \big[a_0^7 g_v^2 b^7+7 a_0^6 g_v^2 r_p b^6+21 a_0^5 g_v^2 r_p^2 b^5+a_0^4 \left(g_c^2-16 g_v g_c+35 g_v^2\right) r_p^3 b^4+7 a_0^3 \left(g_c^2-6 g_v g_c+5 g_v^2\right) r_p^4 b^3
    \nonumber \\
    &+a_0^2 \left(17 g_c^2-38 g_v g_c+21 g_v^2\right) r_p^5 b^2+7 a_0 (g_c-g_v)^2 r_p^6 b+(g_c-g_v)^2 r_p^7\big]
    \Bigg)
\end{align}

\end{widetext}

\section{Estimate of the Computational cost}
\label{app:comp_cost}
{
To provide an idea of the computational cost associated with our method, we take the calculation with the $8\times8\times8$ grid as an example, which is equivalent to a supercell containing 1,024 atoms. 
The bottleneck is the BSE calculation, where 512 independent BSE calculations corresponding to all center-of-mass exciton momenta need to be performed. 
For these calculations we use Frontera supercomputer at Texas Advanced Computing Center. 
Each BSE calculation takes 30 minutes to complete on Intel Xeon Platinum 8280 ``Cascade Lake'' nodes, each supporting 56 compute cores. 
The maximum memory required for each calculation is $\approx$120~MB, and $\approx$50~MB is used to store the BSE eigenvalues and eigenvectors on disk.
In total, 512 node hours are needed to complete all the BSE calculations, and $\approx$25~GB storage space is needed to store the BSE eigenvectors and eigenvalues.
Performing explicit BSE calculations for a 1,024-atom supercell would be computationally prohibitive, since these calculations scale as $O(N^4)$~\cite{giustino2014materials,marsili2017large}.
Importantly, the present approach enables calculations of excitonic polarons extending over tens to hundreds of crystal unit cells.
}

\clearpage
\newpage
\bibliography{literature}

\end{document}